# Detección de comunidades en redes: Algoritmos y aplicaciones

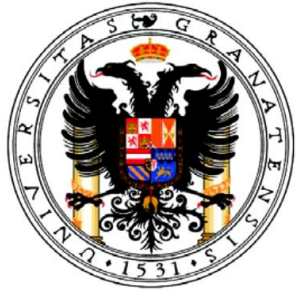

**Julio Omar Ancízar Palacio Niño**

# Detección de comunidades en redes: Algoritmos y aplicaciones

**Julio Omar Ancízar Palacio Niño**

Trabajo de fin de máster presentado como requisito parcial para optar al título de:
**Magister en Soft Computing y Sistemas Inteligentes**

Director:
Ph.D. Fernando Berzal Galiano

Línea de Investigación:
Minería de datos, detección de comunidades

Universidad de Granada
E.T.S. de Ingeniería Informática y de Telecomunicaciones
Departamento de Ciencias de la Computación e Inteligencia Artificial
Granada, España
2013

*A mi familia*

*El modo de dar una vez en el clavo es dar cien veces en la herradura.*

*Miguel de Unamuno*

# Agradecimientos

# Resumen

El presente trabajo de fin de máster tiene como objetivo la realización de un análisis de los métodos de detección de comunidades en redes. Como parte inicial se realizó un estudio de las características principales de la teoría de grafos y las comunidades, así como medidas comunes en este problema. Posteriormente, se realizó una revisión de los principales métodos de detección de comunidades, elaborando una clasificación, teniendo en cuenta sus características y complejidad computacional, para la detección de las fortalezas y debilidades en los métodos, así como también trabajos posteriores. Luego, se estudio el problema de la calificación de un método de agrupamiento, esto con el fin de evaluar la calidad de las comunidades detectadas, analizando diferentes medidas. Por último se elaboraron las conclusiones así como las posibles líneas de trabajo que se pueden derivar.

**Palabras clave:** *clique*, *cluster*, *clustering*, complejidad computacional, comunidad, detección de comunidades, grafo, método, métrica, redes, solapamiento, subgrafo

# Abstract

This master's thesis work has the objective of performing an analysis of the methods for detecting communities in networks. As an initial part, I study of the main features of graph theory and communities, as well as common measures in this problem. Subsequently, I was performed a review of the main methods of detecting communities, developing a classification, taking into account its characteristics and computational complexity for the detection of strengths and weaknesses in the methods, as well as later works. Then, study the problem of classification of a clustering method, this in order to evaluate the quality of the communities detected by analyzing different measures. Finally conclusions are elaborated and possible lines of work that can be derived.

# Contenido

# Lista de figuras

# Lista de tablas

# 1. Introducción

## 1.1 Motivación

El estudio de la detección de comunidades es un área interdisciplinar, cuyos inicios se remontan a principio del siglo XX en investigaciones realizadas por varias ramas de las ciencias humanas, como la sociología, antropología, psicología, acuñando el término de sociometría. Su objetivo es el estudio del comportamiento de las sociedades y las relaciones entre sus individuos por medio de modelos de redes [45].

Gracias a su practicidad en el modelamiento de sistemas, su implementación se extendió por varias ramas de la ciencia como la física, biología, ciencias sociales y las ciencias de la computación, lo que ha permitido el surgimiento de varios métodos de detección de comunidades desarrollados desde varios puntos de vista [19].

El trabajo propuesto consiste en el estudio y el análisis de los métodos de evaluación que se han propuesto para evaluar los resultados obtenidos en dichas técnicas. En particular el trabajo se centrara en las denominadas técnicas de detección de comunidades (análogas a las técnicas de *clustering* tradicionales en otros ámbitos de la minería de datos).

Entre las técnicas tradicionales de detección de comunidades se encuentran, por ejemplo:

- Técnicas de tipo jerárquico, que puede utilizar medidas de centralidad o importancia asociada a enlaces concretos (por ejemplo *betweenness*) al estilo del *single-link clustering*, medidas globales de modularidad que intentan detectar zonas de la red en las que hay mas enlaces de los que se esperaría en una red meramente aleatoria u otras medidas de similitud que permitan decidir cómo realizar el agrupamiento.

- Técnicas basadas en la identificación de *cliques*, en las que se identifican conjuntos de nodos muy conectados entre sí y se van fusionando conjuntos adyacentes.

El proceso de agrupamiento por lo general no es realizado por los humanos, por lo que el proceso de agrupamiento es útil para descubrir posibles agrupaciones y comportamientos en los datos analizados [29]

La detección de comunidades en redes tiende a utilizarse en problemas aplicados a las redes donde pueden surgir métodos de solución particulares, como por ejemplo el adecuado enrutamiento en redes inalámbricas [54], análisis de sistemas de transporte [88].

El principal problema de estas técnicas es su eficiencia, al menos cuadrática en el caso de las técnicas jerárquicas y basadas en la resolución aproximada de problemas NP en el caso de la identificación de *cliques*. Por tanto, resulta de especial interés tanto teórico como practico el desarrollo de variantes de estas técnicas que sean verdaderamente escalables.

Trabajos realizados con respecto a los métodos de detección de comunidades, como el realizado por Fortunato[19], permiten evidenciar la multiplicidad de técnicas para la solución de un mismo problema, lo cual evidencia la falta de un consenso en la resolución del problema de detección de comunidades, además de la multiplicidad de métodos para medir la calidad de las mismas.

Otro punto a tener en cuenta en esta área es la falta de delimitación del problema, lo que provoca el desarrollo de métodos que tienen funcionamientos "óptimos" bajo ciertos parámetros, lo cual hace que su aplicación práctica sea restringida a la solución de problemas específicos, además no hay consenso en el empleo de métricas estándares y métodos formales de evaluación.

## 1.2 Objetivos

Los objetivos del presente trabajo de fin de máster son:

- Hacer una revisión bibliográfica de los conceptos generales de la detección de comunidades en redes.

- Revisar y analizar los métodos no supervisados de detección de comunidades en redes, agruparlos en una taxonomía adecuada; establecer características básicas, fortalezas, debilidades así como investigaciones actuales.

- Revisar y analizar los métodos para evaluar la calidad de los métodos de detección de comunidades, clasificarlos adecuadamente resaltando sus fortalezas y debilidades.

- Basado en los métodos sobre detección de comunidades, detectar las fortalezas y debilidades de los métodos analizados, teniendo encuentra sus características intrínsecas y la complejidad computacional.

## 1.3 Organización del trabajo

Este trabajo está organizado de la siguiente manera:

En el capítulo 2 se realiza una revisión de a las bases del problema de detección de comunidades, dividiéndolo en 3 partes: primero se establece una revisión de la teoría base para familiarizarse con la teoría de grafos para representación de redes y sus características principales, vitales para la comprensión de los métodos y métricas empleadas en el trabajo; segundo se revisa el tema correspondiente a los subgrafos y comunidades, el cual es el concepto clave para la detección y acercarse al problema de la definición de una comunidad y las características a tener en cuenta dependiendo del enfoque empleado; en la tercera parte se hace una revisión de las medidas más comunes que se tienen en cuenta para los métodos de detección de comunidades, cuyo objetivo es cuantificar al *cluster*.

En el capítulo 3 se realiza una revisión de varias técnicas de detección de comunidades, basadas en métodos no supervisados. Los métodos se clasifican en base a sus características, medidas empleadas, entre otras. Sobre cada método se describe su funcionamiento, métricas empleadas y complejidad computacional. Finalmente se presenta los métodos de detección de comunidades agrupados en un cuadro comparativo, resaltando sus características principales.

En el capítulo 4 se describe las métricas empleadas para evaluar la calidad de los resultados de un proceso de detección de comunidades, clasificándolos en dos conjuntos principales, aquellos que no tienen información adicional a la contenida en la red, y aquellas métricas que emplean información adicional como resultados de otros métodos sobre la misma red o una solución optima. Por último se realiza un resumen de las características principales que se tiene en cuenta por parte de las métricas para la evaluación de un *cluster* como "bueno", debido a que cada métrica enfatiza sobre algunas características, pero tiene a fallar sobre condiciones especificas que se pueden presentar.

Por último se muestran las conclusiones y líneas de investigación futuras del presente trabajo; resaltando la importancia de la detección de comunidades, métodos empleados y las métricas adecuadas para su evaluación.

# 2. Redes y comunidades

Una red es una representación de un sistema que modela los enlaces entre sus elementos. También puede incluir información adicional en forma de etiquetas y pesos. Como cualquier modelo, la posible pérdida de información conlleva desventajas, pero la reducción de un sistema a una red puede conducir a ventajas en su proceso de análisis[51].

Varias áreas de interés como la física, la biología o las ciencias humanas emplean el modelado de redes para su análisis. Por ejemplo, las redes sociales, que en un principio fueron estudiadas por las ciencias humanas, hoy en día también son estudiadas por la estadística y las ciencias de la computación, entre otras [51].

Una red social es una red de personas o de grupos de personas. Estas personas o grupos se representan como los nodos de la red y las aristas representan las conexiones entre ellos, como las amistades entre personas o los vínculos entre grupos de personas. Varias de las herramientas utilizadas para el análisis de estas redes corresponden al área de la estadística [51].

**Figura 2-1:** Red de un club de karate (Zachary)[19].

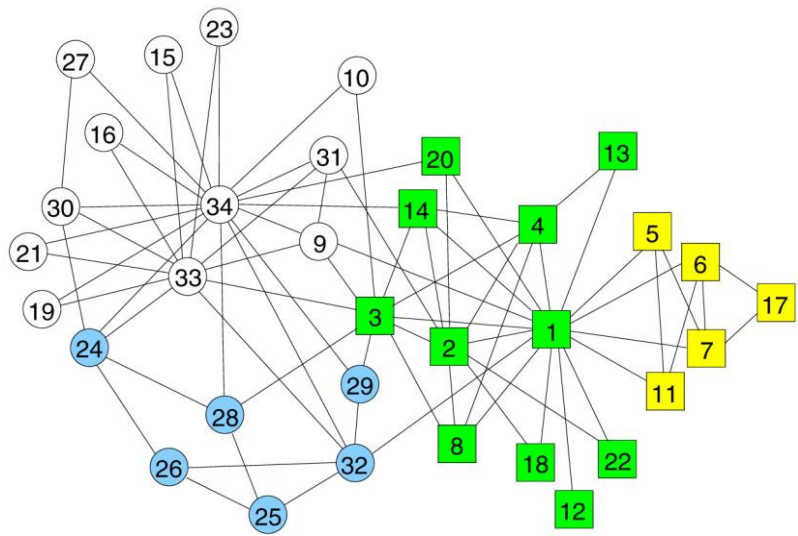

Un ejemplo conocido de una red social es el club de karate de Wayne Zachary (Figura 2-1) donde la red modela las amistades existentes entre los miembros del club de karate en una universidad norteamericana. La construcción de los vínculos reflejados en la red se realizó por observación en el comportamiento del grupo. [51]

Como se indicó anteriormente, aparte de las redes sociales, existen otros tipos de modelos de redes de gran interés, como [51]:

- Redes de transporte (carreteras, comunicaciones, energía).

- Redes de información (WWW, e-mail, citas bibliográficas).

- Redes biológicas (bioquímicas, neuronales, ecológicas).

En esté capitulo se expondrán los conceptos preliminares correspondientes a la representación de datos por medio de grafos, empezando por unas definiciones y características de las redes y comunidades, que se puede considerar una comunidad visto desde varios enfoques, y por último se revisará las medidas más usuales utilizadas como fundamento para la elaboración de los métodos de detección de comunidades, fortaleciendo alguna de las características.

## 2.1 Definiciones y características

Una **red** es una forma de representar las relaciones entre los elementos de una colección. Se compone de un conjunto de objetos, llamados **nodos** o **vértices**, los cuales se encuentran conectados por medio de enlaces, llamados **aristas** o **arcos**. Si dos nodos se encuentran conectados por medio de una arista, se denominan nodos vecinos [16].

En matemáticas, el concepto más básico de red se conoce también como grafo [67]. A continuación, se presentará la definición formal de grafo y sus características más relevantes:

### 2.1.1 Definición

Un **grafo no dirigido**[1] es un par $G = (V, E)$, donde $V$ es un conjunto (finito) de vértices o nodos, $E$ es un conjunto de pares no ordenados de elementos distintos de $V$. A estos pares se les llama **aristas** [67].

Existen situaciones en las cuales es necesario modelar más de una arista por cada par de nodos. Esta situación se conoce como **multígrafo**. Al igual que un grafo, un multígrafo consta de nodos y aristas, pero con la diferencia de que no existe restricción en la existencia de aristas múltiples entre pares de nodos [67].

Cuando existe una arista para conectar un nodo consigo mismo, esta arista se conoce como **bucle**, pero con un grafo simple o un multígrafo no es posible modelarlo. Para esto se utiliza un **pseudografo** [67].

Por el contrario, un **grafo dirigido** $G = (V, E)$ consiste en un conjunto no vacío $V$ de nodos y de un conjunto $E$ de aristas dirigidas. Esta dirección es asociada con un par ordenado de nodos. Una **arista dirigida** es asociada con un par ordenado $(u, v)$ donde el nodo $u$ es el inicio y $v$ es el final de la arista.

Por lo general, un grafo dirigido se representa de la siguiente manera $\vec{G} = (V, \vec{E})$.

**Figura 2-2:** Ejemplo de grafo no dirigido y de grafo dirigido [16].

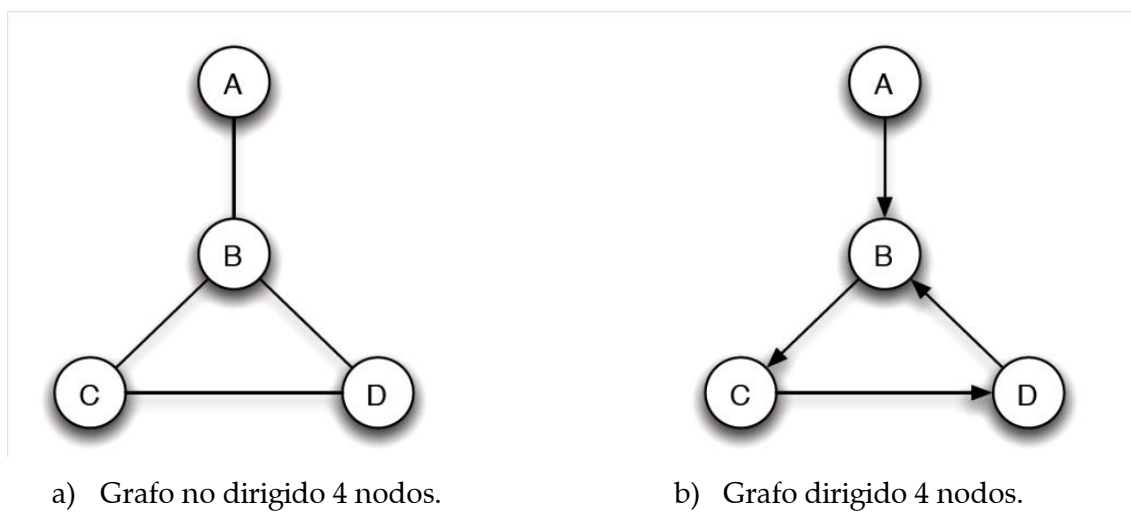

a) Grafo no dirigido 4 nodos.    b) Grafo dirigido 4 nodos.

---

[1] Para facilitar la legibilidad del texto se considerará en adelante como, simplemente, **grafo**.

## 2.1.2 Adyacencia

Sea $G = (V, E)$ un grafo. Un nodo $u \in V$ es **adyacente** o **vecino** a un nodo $v \in V$ si y sólo si hay un arista entre ellos; es decir, si y sólo si $(u, v) \in E$. El conjunto de vecinos de $i$ es [4]:

$$N(u) = \{v \in V | (u, v) \in E\}$$

Si $e = \{u, v\}$, se dice que la arista $e$ es **incidente** en los nodos $u$ y $v$. También se dice que la arista $e$ **conecta** $u$ y $v$.

## 2.1.3 Peso

Existen casos en los cuales las redes no sólo representan conexiones simples de tipo booleano. Algunas veces es necesario modelar un peso, fuerza o valor asociado a la conexión. Por ejemplo en un modelo de una red de comunicaciones, los nodos representan terminales y enrutadores, mientras que las aristas representan las conexiones y un valor asociado a cada arista puede representar el ancho de banda del que se dispone en esa conexión [51].

El **peso** de una arista representa un valor intrínseco a ella. Cuando en un grafo se modelan los pesos, este se denomina **grafo ponderado**.

## 2.1.4 Grado

El **grado** o **conectividad** de un nodo de un grafo es el número de aristas incidentes en él, cada arista contribuye con dos unidades al grado de los nodos que conecta y por lo general se denota $k_i$ o $\deg(i)$ [67]. Esta medida es de gran importancia en varios métodos de detección de comunidades, ya que permite comprender de una manera la relación entre los nodos y el número de vecinos adyacentes al nodo.

De manera formal el grado de un nodo se puede expresar como [4]:

$$\deg(i) = |N(i)|$$

El conjunto $N(i)$ es algunas veces llamado la **frontera o límite** de $v$. El límite de $v$ junto con $v$ es llamado el **cierre** de $v$ [4].

$$\text{cierre}(v) = N(v) \cup \{v\}$$

A los nodos de grado cero, $\deg(i) = 0$, se les llama **aislados. S**i $\deg(i) = 1$, se les denomina nodos **colgantes** u **hojas.**

Existe una relación directa entre el número de aristas $e$ y el grado de cada uno de los nodos $\deg(i)$ en un grafo $G = (V, E)$. Esta propiedad se conoce como el **teorema de los apretones de manos:**

$$2e = \sum_{v=V} \deg(v)$$

Formalmente, el grado en un nodo en un grafo ponderado se define como [81]:

$$\deg(i) = \sum_{j=1}^{n} w_{ij}$$

Donde $w_{ij}$ es el peso de la arista del par de nodos $(i, j)$.

La **matriz de grados** $D$ se define como la matriz con los grados $\deg(1), \ldots, \deg(n)$ en la diagonal, [81]

$$D = \begin{bmatrix} \deg(1) & & \\ & \ddots & \\ & & \deg(n) \end{bmatrix}$$

$$D(i,j) = \begin{cases} 0 \text{ si } i \neq j \\ \deg(v_i) \text{ si } i = j \end{cases} \Rightarrow D = \text{diag}(\deg(1), \ldots, \deg(n))$$

### 2.1.5 Representación

Existen varias formas de representar los grafos, como modelos gráficos, listados, tablas o matrices. A continuación, se explicarán brevemente las representaciones matriciales, útiles cuando hay que analizar grafos matemáticamente.

#### 2.1.5.1 Matriz de adyacencia

La **matriz de adyacencia** permite representar, para un grafo no dirigido $G = (V, E)$ con un conjunto de nodos $V = \{v_1, \ldots, v_n\}$, las parejas de nodos que se encuentran interconectados. La matriz $A = [a_{ij}]$ es de tamaño $n \times n$, donde $n$ es el número total de nodos [81].

La matriz de adyacencia se define como [51]:

$$a_{ij} = \begin{cases} 1, & \text{si existe una arista entre el nodo } i \text{ y el nodo } j \\ 0, & \text{en caso contrario} \end{cases}$$

De la misma manera, un grafo ponderado se puede representar mediante una matriz de adyacencia [51].

$$a_{ij} = \begin{cases} w, & \text{si existe una arista con peso } w \text{ entre el nodo } i \text{ y el nodo } j \\ 0, & \text{en caso contrario} \end{cases}$$

#### 2.1.5.2 Matriz de incidencia

La **matriz de incidencia** permite representar un grafo no dirigido $G = (V, E)$ con un conjunto de nodos $V = \{v_1, \dots, v_n\}$ y un conjunto de aristas $E = \{e_1, \dots, e_m\}$. La matriz de incidencia es de tamaño $m \times n$ y se define como $M = [m_{ij}]$ [67]:

$$m_{ij} = \begin{cases} 1, & \text{si la arista } e_j \text{ es incidente con } v_i \\ 0, & \text{en caso contrario} \end{cases}$$

### 2.1.6 Caminos

Si un par de nodos $(i, j)$ no adyacentes pueden ser conectados por medio de una secuencia de $m$ aristas $(i, l_1), (l_1, l_2), \dots, (l_{m-1}, j)$; este conjunto de aristas describe un **camino** entre el nodo $i$ y el nodo $j$, donde $m$ es la **longitud del camino** [9].

Se puede decir que dos nodos se encuentran **conectados** si existe al menos un camino entre ellos.

Si todos los nodos y aristas que componen un camino entre dos nodos $i$ y $j$ no se repiten, el camino se conoce comúnmente como **camino simple** o *path* [9].

Un **bucle** se define como un camino donde el nodo inicial y final es el mismo nodo y se pasa una única vez por cada uno de los $m - 1$ nodos intermedios [9].

## 2.2 Subgrafos

Sea $G = (V, E)$ un grafo no dirigido. $G_X = (X, E_X)$ es un **subgrafo** de $G$ (inducido por $X$) si y sólo si $X \subseteq V$ y $E_X \leq (X \times X) \cap E$; es decir, si contiene un subconjunto de nodos de $G$ y las aristas correspondientes [4].

Recíprocamente, el grafo $G$ es un **supergrafo** de $G_X$ si $G_X$ es un **subgrafo** de $G$. Algunas veces, la condición de subgrafo hace referencia sólo a la condición de $E_X \subseteq E$ y un **subgrafo inducido** será aquel que cumpla la definición de subgrafo [8].

Un grafo $G = (V, E)$ es un **grafo completo** si incluye todas las aristas posibles. Formalmente el grafo completo $K_n$ de nodos $n$ se define como:

$$K_n = V \times V - \{(v, v) | v \in V\}$$

**Figura 2-3:** Ejemplo de subgrafo [79].

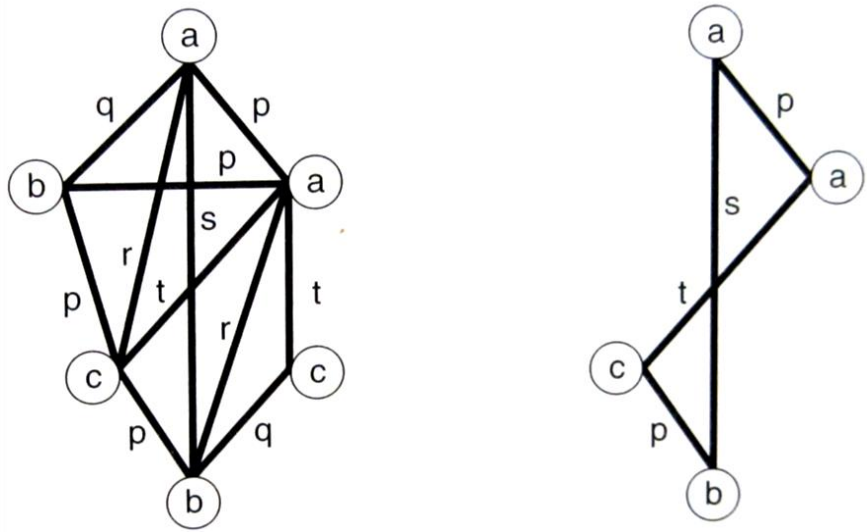

Un *clique* es un conjunto de nodos que se encuentran todos interconectados. Por lo tanto, cumplen con la definición de subgrafo completo. Un *clique* es llamado **máximo** si y sólo si no puede ampliarse mediante la inclusión de un vértice adyacente [49].

Un ***k-clique*** es un grupo de nodos conectados, en el que $k$ es la máxima longitud del camino entre un par de nodos del *k-clique*. De manera más formal, se define como un subconjunto de nodos $C$ tal que para todo $i, j \in C$, la distancia $d(i, j) \leq k$ [51].

Por lo tanto, un *clique* es también un *1-clique*, ya que la distancia máxima entre cualquier par de nodos es directa. Un *2-clique* es un grafo donde el camino que conecta cualquier par de nodos tiene una longitud $k \leq 2$. Este tipo de agrupamientos se puede asociar a los "amigos de amigos" en las redes sociales. En la **Figura 2-4** se pueden ver gráficamente ejemplos de *k-cliques* con $k$ de 1 a 3 [51].

**Figura 2-4:** Ejemplos con cuatro nodos de *1-clique, 2-clique* y *3-clique* [51].

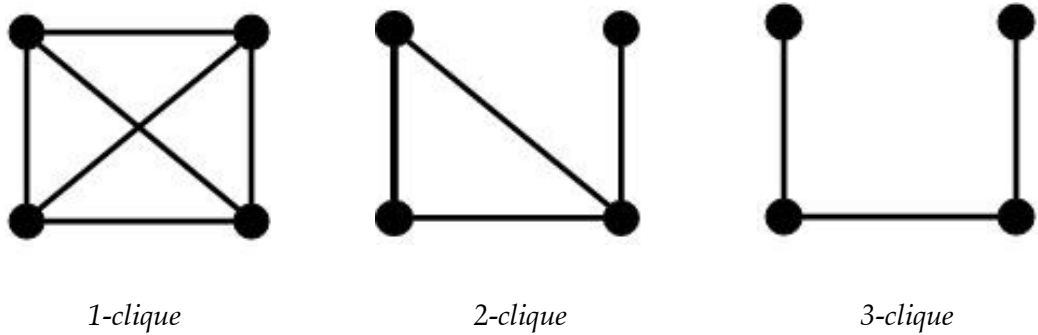

*1-clique*  *2-clique*  *3-clique*

Un **k-plex** de $n$ nodos es un subconjunto máximo de $n$ nodos dentro de una red, de tal manera que cada nodo está conectado a, por lo menos, $n - k$ nodos de los demás nodos del *k-plex*. Si $k = 1$, conforma un *clique*; si $k = 2$, entonces cada nodo debe estar conectado a todos o a todos menos un nodo y así sucesivamente. Además, los *k-plexes* pueden solaparse si comparten nodos [51].

Un **k-core** es un subconjunto máximo de $n$ nodos de tal manera que cada nodo está conectado a por lo menos otros $k$ nodos en el subconjunto. A diferencia de los *k-plexes*, los *k-cores* no pueden formar grupos solapados [51]. Adicionalmente, un *k-core* tiene la propiedad de contener *k-cliques* y *k-plexes*, donde un *k-plex* de tamaño $s$ forman parte de un $(s - k)$-*core* [59].

Uno de los grandes inconvenientes de resolver problemas por medio de la identificación de *cliques* es su complejidad computacional: la búsqueda de *cliques* en un grafo es un problema NP-completo [19].

Se ha demostrado que resolver el problema de *k-clique* máximo es un problema NP-difícil para cualquier entero positivo $k$. Existen varias formas de afrontar este tipo de problemas; una es por medio de reducir el problema a la máxima potencia $k$ del grafo ($G^k$), donde $G^k$ es la reducción de un grafo $G$, de tal forma que la longitud máxima entre un nodo y otro está dada por $k$. En algunos casos, un problema de detección de *k-* *cliques* puede ser más difícil que la detección de *cliques*, aunque puede presentar buenos resultados con grafos dispersos [59].

Los problemas de detección de *k-plexes* también presentan una complejidad NP-difícil, por lo que se hace necesario la implementación de heurísticas para obtener soluciones aproximadas. También se emplean enfoques de programación lineal entera [59].

Por otra parte, la detección de *k-core* es un problema polinómico, por lo que un algoritmo voraz es capaz de obtener una buena respuesta. Además, la detección de *k-cores* se implementa como una etapa de preprocesamiento en la detección de *cliques*, *k-cliques* y *k-plexes* [59].

## 2.3 Comunidades

Uno de los problemas fundamentales de cara al desarrollo de técnicas de detección de comunidades es definir de manera cuantitativa una comunidad en un grafo. Debido a que no existe una definición universal para este problema, la caracterización de las comunidades debe realizarse dependiendo del problema analizado. De manera intuitiva, podríamos definir una comunidad como aquel conjunto de nodos interconectados por una mayor cantidad de aristas que los demás nodos[19].

La definición del término "comunidad" no es una definición absoluta, depende desde el punto de vista desde el cual se estudia. Desde un punto de vista social, una comunidad se define como un conjunto de personas con intereses o actividades en común que formen parte de una sola red [58]. Una definición es la propuesta por Newman y Girvan es "una comunidad es la división de los nodos de una red en grupos dentro de los cuales las conexiones son densas, pero entre ellos son escasas"[52], esta definición se puede ver en la **Figura 2-5**.

Desde un punto de vista matemático, basado en la definición anteriormente vista, Papadopoulos define una **comunidad** $C$ como un subgrafo de un grafo $G = (V, E)$, el cual comprende un conjunto de nodos que se encuentran interconectados más densamente que el resto de nodos, debido a que comparten un interés común[65].

**Figura 2-5:** Ejemplo de comunidad [52].

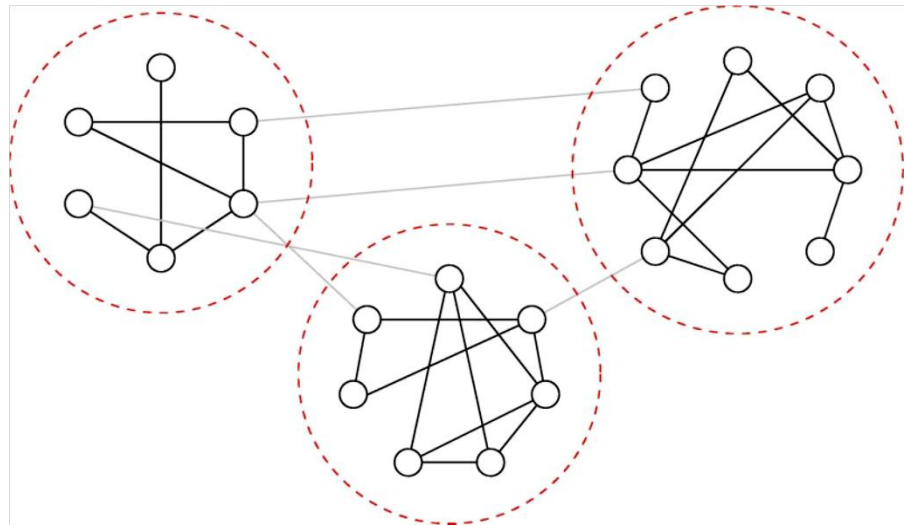

En la **Figura 2-5** se describe gráficamente el concepto de Girvan y Newman, así como también el de Papadopoulos donde en el grafo se pueden identificar claramente tres comunidades, las cuales tienen una alta interconexión entre sus nodos y las aristas que conectan a las comunidades se encuentran en una menor cantidad. Varios de los métodos de detección de comunidades se centran en resaltar estas características.

Trabajos realizados por Reichardt y Bornholdt [65] a partir de las definiciones de comunidad plantadas por Newman [49], definen de manera formal una comunidad como un conjunto de nodos para el que se verifican las siguientes desigualdades:

$$\frac{2m}{n(n-1)} > \frac{2M}{N(N-1)} > \frac{m_{nN}}{n(N-n)}$$

donde $m$ es el número de aristas entre los $n$ nodos de la comunidad y $m_{nN}$ es el número de aristas que conectan a los restantes $N - n$ nodos de la red. En la cual, de manera general, una comunidad debe presentar una densidad mayor con respecto a la densidad promedio del grafo, y esta a su vez debe ser superior a la densidad externa a la comunidad [65].

Las comunidades **explícitas** son aquellas que se construyen a partir de la intervención humana, es el humano quien decide los enlaces entre los nodos de la red. Ejemplos de este tipo de comunidades son Facebook, Google+, LinkedIn o Flick [58].

Las comunidades **implícitas** son aquellas donde se supone la existencia de una comunidad en un conjunto de datos y el sistema es el encargado de "descubrir" la existencia de esas comunidades. La investigación en el desarrollo de técnicas que permitan descubrir de manera automática las comunidades ofrece ventajas al no requerir la intervención humana en su elaboración[58].

A partir de la idea descrita anteriormente, donde una comunidad se caracteriza por una mayor cantidad de conexiones entre sus nodos con respecto a los de fuera, varias definiciones se han ideado desde el campo de las ciencias humanas y, de manera más reciente, desde la física y las ciencias de la computación, distinguiéndose tres tipos de definición, local, global y basados en la similitud de nodos [19].

## 2.3.1 Definición local

Desde el punto de vista local, una comunidad es un conjunto con mayor conexión entre los nodos dentro que entre los de fuera de la comunidad, hasta el punto que se pueden analizar de manera independiente del resto de la red. El estudio de las comunidades basadas en este concepto centran sus estudios en el subgrafo y los nodos vecinos de manera inmediata, pero dejando de lado el resto de la red [19].

La detección de comunidades locales en las redes se fundamenta en la búsqueda de conjuntos de nodos altamente conectados como son los *cliques, n-cliques, n-clans, k-plexes* y *k-cores* [58].

## 2.3.2 Definición global

El análisis de redes también se puede realizar sobre todo su grafo en conjunto. Estos casos se presentan cuando las comunidades no pueden "aislarse" del resto de la red, ya que forman parte fundamental de su estructura. La idea fundamental parte de la comparación de la red analizada con el modelo de una red aleatoria. Por definición, una red aleatoria no tiene una estructura de comunidad en su interior. Si se presentan diferencias entre la red analizada y la aleatoria, se puede inferir que existe algún tipo de comunidad [19].

Para este análisis, se pueden emplear varios tipos de medidas, que permite realizar estas comparaciones, como lo son la modularidad [52]; conductancia [32]; el corte normalizado [72], entre otras.

### 2.3.3 Definición basada en similitud

Es de suponer que los integrantes de una comunidad tienen características similares. Esta similitud se puede calcular mediante alguna medida ya sea que los nodos estén conectados por una arista o no. Entre las medidas más comunes de similitud podemos nombrar la distancia euclídea, la distancia Manhattan o la similitud del coseno, entre otras muchas. [19].

Pero las comunidades no se definen únicamente por las similitudes entre nodos. También se emplean otro tipo de métricas, como el número de caminos independientes entre dos nodos o el flujo máximo que puede transitar entre los dos nodos. Este flujo máximo se puede determinar con ayuda del teorema del flujo máximo / corte mínimo [17], en tiempo polinómico [19].

## 2.4 Medidas en redes y comunidades

Una de las áreas importantes en el estudio de las redes y comunidades es determinar que características se deben resaltar de los grupos a encontrar. Estas características pueden ser de centralidad, transitividad, entre otras. Muchas de estas medidas se originan en los primeros estudios sobre redes sociales, elaborados por sociólogos y demás profesiones afines a las ciencias humanas, que hoy en día se han complementado con otros enfoques como la física, las ciencias de la computación o la biología.

### 2.4.1 Distancia

En un grafo $G = (V, E)$ la **distancia** entre dos nodos $i$ y $j$ se define como el camino más corto existente entre ellos. Si $i = j$, entonces $d(i, j) = 0$. Si no existe un camino posible entre $i$ y $j$, entonces $d(i, j) = \infty$ [9].

Si el grafo es ponderado la distancia se calcula como la suma de los pesos asociados a las aristas del camino [9]:

$$d(i,j) = \sum w_{i,j}$$

Un **camino geodésico** es el camino más corto que puede existir entre un par de nodos que se encuentren conectados por medio de un camino [9].

La **distancia media o camino medio** mide la distancia promedio que existe entre un par de nodos $i$ y $j$ que se comunican por el camino geodésico, de manera formal se define como [9]:

$$d_m(G) = \frac{1}{n(n-1)} \sum_{i,j} d(i,j)$$

El **diámetro** en una red es la máxima distancia que puede existir entre dos nodos $i$ y $j$ de entre todos los nodos del grafo que estén unidos por un camino geodésico. Esta medida es de gran importancia para establecer el tamaño de la red [9].

$$\text{Diámetro}(G) = \max\{d(i,j)\} \, \forall \, i,j \in V$$

Para calcular su valor, se deben calcular todos los caminos geodésicos para cada par de nodos existentes en la red y, de ellos, escoger el de mayor longitud.

## 2.4.2 Densidad

La **densidad** $\delta(G)$ es la relación entre el número de aristas presentes en el grafo ($m$) y el número máximo de aristas que puede tener el grafo, cuyo valor se calcula a partir del número de nodos presentes ($n$) [55].

De manera formal, se expresa como:

$$\delta(G) = \frac{m}{n(n-1)}$$

En un grafo completo, la densidad presente es máxima, $\delta(G) = 1$.

La **densidad intra-cluster** $\delta_{int}(C)$ es la relación entre en número de aristas internas y el número aristas que podrían existir dentro de un cluster. [19].

$$\delta_{int}(C) = \frac{\#\text{aristas internas de } C}{\frac{n_c(n_c-1)}{2}}$$

Recíprocamente, se define la **densidad inter-cluster** $\delta_{ext}(C)$ como la relación entre el número de aristas que conectan el grupo con el resto del grafo y el número de todas las posibles aristas. [19].

$$\delta_{ext}(C) = \frac{\#\text{ aristas externas de } C}{n_c(n-n_c)}$$

### 2.4.3 Centralidad

Las medidas de centralidad ayudan a evaluar los nodos en busca de aquéllos que sean más importantes o céntricos en una agrupación [51]. El concepto de la centralidad se inicia en los estudios de comportamiento humano y sus relaciones; basándose en el uso de conceptos matemáticos, estadísticos y de teoría de grafos [21].

Hay diferentes tipos de medidas para la centralidad, a continuación se explicarán algunos de ellos: centralidad por grado *(degree centrality)*, centralidad por vector propio *eigenvector centrality)*, centralidad por cercanía *(closeness centrality)* y centralidad por intermediación *(betweenness centrality)*.

#### 2.4.3.1 Centralidad por grado (*Degree centrality*)

Como se vio anteriormente, el **grado** de un nodo en un grafo es el número de aristas conectadas a él. Se denota como $k_i$ donde $i$ representa el nodo. De manera formal, se expresa de la siguiente forma a partir de la matriz de adyacencia del grafo:

$$k_i = \sum_{j=1}^{n} a_{ij}$$

El **grado medio** de un grafo se determina como:

$$c = \frac{1}{n}\sum_{i=1}^{n} k_i$$

Sea un subgrafo $C$ de un grafo $G$, donde $|C| = n_c$ y $|G| = n$ nodos, respectivamente. El **grado interno** de un nodo $v \in C$ es $k_v^{int}$, donde el numero de aristas conectados en $v$ con otros nodos de $C$. Recíprocamente, se aplica este concepto al **grado externo** $k_v^{ext}$ como el número de aristas que conectan nodos de $C$ con nodos del resto del grafo ($v \notin C$) [19].

Si $k_v^{ext} = 0$, los nodos son vecinos únicamente de otros nodos de $C$, indicando que el nodo $v$ pertenece al *cluster* $C$. Si $k_v^{int} = 0$, el nodo no pertenecerá a $C$, si bien puede pertenecer a otro *cluster* [19].

El **grado interno del subgrupo** $C$, $k_C^{int}$ es la suma de los grados internos de los nodos pertenecientes al clúster, de la misma manera el **grado externo del subgrupo** $C$, $k_C^{ext}$ es la suma de los grados externos de los vértices [19].

El **grado total del subgrupo** $C$ es la suma de los grados de los nodos de $C$ [19].

$$k_C = k_C^{int} + k_C^{ext}$$

### 2.4.3.2 Centralidad por vector propio *(Eigenvector Centrality)*

En la centralidad por grado se considera la importancia de un nodo con respecto a su entorno por el número de conexiones. En este caso, se considera la importancia de que el nodo tenga conexiones con otros nodos de igual importancia, siendo esta la base para la centralidad por vector propio *(eigenvector centrality)* [51].

Esta medida inspiró el desarrollo del algoritmo de PageRank el cual sirve de base para el motor de búsqueda de Google [51].

Formalmente la **centralidad por vector propio** se define de la siguiente manera [51]:

$$C_e(i) = \frac{1}{K_1} \sum_j A_{ij} x_j$$

Donde $A_{ij}$ es la matriz de adyacencia, $K_1$ es una constante [51].

Con esta medida, un nodo adquiere "importancia" por medio de dos caminos: tener muchas conexiones con otros miembros de la comunidad o tener conexiones con otros nodos "importantes", aunque sean pocos [51].

Una de las limitantes de esta medida es el análisis sobre redes dirigidas [51].

### 2.4.3.3 Centralidad por cercanía *(Closeness Centrality)*

La centralidad por cercanía consiste en calcular las distancias que existen entre el nodo analizado y los demás nodos. Cuanto más grande es la distancia, esta representa menor importancia y, recíprocamente, las distancias menores indican mayores distancias [51].

De manera formal, se define como [51]:

$$C_c(i) = \frac{n}{\sum_{j=1}^{n} d_{ij}}$$

Donde $C_c(i)$ es la centralidad por cercanía del nodo $i$, $n$ es el número de nodos y $d_{ij}$ la distancia que hay desde el nodo $i$ hasta el nodo $j$. [51]

Uno de los inconvenientes que presenta este tipo de centralidad es que solo puede ser utilizado en redes donde exista un camino entre cada par de nodos existentes en la red. Otro inconveniente consiste en su variable complejidad computacional, que puede variar de $O(mn)$ hasta $O(n^4)$ [51].

### 2.4.3.4 Centralidad por intermediación *(Betweenness Centrality)*

La medida de intermediación o de *betweenness* se define como una medida para identificar las aristas que conectan comunidades, otorgando valores altos para aquellas aristas que conectan nodos fuera de la comunidad y penalizando aquellas aristas que se encuentran dentro de la comunidad [52].

Para calcular la medida de *betweenness*, Girvan y Newman exponen 3 métodos [52]:

- **Camino más corto (*Shortest - path*)**

Se basa en calcular el camino más corto entre un par de nodos, para todo el conjunto de nodos existentes en la red, y contar cuantos "caminos" pasan por cada arista, tomando como criterio de elección las aristas con mayor importancia [52].

De manera formal, la medida de *betweenness* se define para cada arista $e \in E$ como[52]:

$$C_B(e) = \sum_{u,v \in V} \frac{g_e(u,v)}{g(u,v)},$$

Donde $g(u,v)$ es el número total de caminos mínimos entre los nodos $u$ y $v$, y $g_e(u,v)$ es el número de caminos mínimos entre los nodos $u$ y $v$ que pasan por el nodo $e$. Valores pequeños de *betweenness* indican que las aristas pertenecen a una misma comunidad y aquellas aristas que conectan nodos de diferentes comunidades tendrán valores más altos [52].

**Figura 2-6:** Cálculo de *betweenness* por el camino más corto [52].

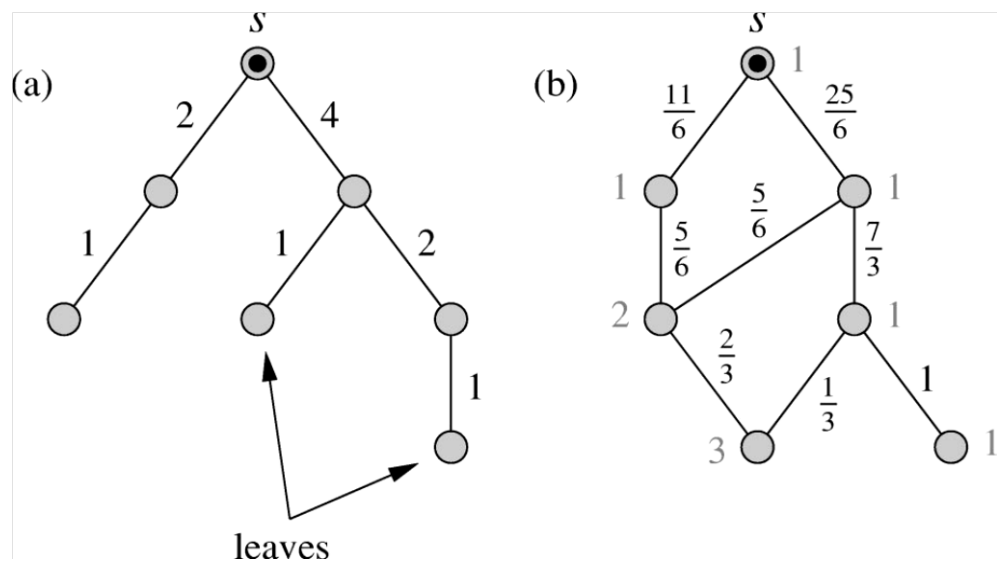

En la **Figura 2-6** se puede apreciar el cálculo del *betweenness* para un árbol (a), donde sólo hay un posible camino desde cualquier nodo al nodo $S$. En este tipo de red es muy simple el procedimiento. Para el caso (b), donde existe más de un camino desde cualquier nodo al nodo $S$, el cálculo es más complejo.

- **Flujo de corriente (*Current - flow*)**

Tanto por el procedimiento por ruta más corta como por el de caminos aleatorios, siempre se está determinando un flujo o una importancia de la arista. Sobre esta premisa se inspira el método del flujo de corriente, en el cual la red se modela como un circuito eléctrico, donde las aristas se consideran resistencias eléctricas, (ver **Figura 2-7**) y donde se determina la corriente que fluye desde un nodo $s$ hasta un nodo $t$ [52].

**Figura 2-7:** ejemplo de una red con resistencias[52].

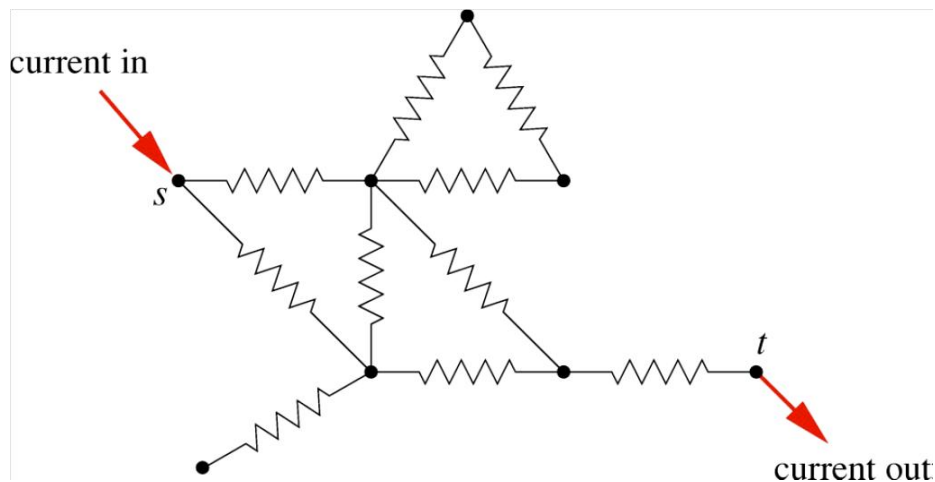

Si se aplica una diferencia de voltaje entre los dos nodos $(s,t)$, cada arista lleva una cierta cantidad de corriente, la cual se puede calcular mediante el uso de las leyes de Kirchhoff, repitiéndose este procedimiento en todos los nodos; como resultado se obtiene es el promedio de la corriente que se transporta entre el par de nodos. Este resultado es homologo a los otros métodos realizados, (ruta mínima, camino aleatorio) [19].

- **Camino aleatorio (*Random - walk*)**

Otro enfoque para calcular la medida de *betweenness* se basa en considerar caminos aleatorios. En este caso, la medida de *betweenness* de una arista está dada de forma aproximada por la frecuencia con la que se pueden generar recorridos entre un par de nodos $(s,t)$.

Se escogen al azar un par de nodos $(s,t)$, el "caminante" empieza en $s$ y se mantiene en movimiento hasta que llega a $t$, donde se detiene. Se calcula la probabilidad de cada arista por la cual se cruza y se realiza el promedio de todos los recorridos posibles entre $s$ y $t$.

## 2.4.4 Coeficiente de agrupamiento o *clustering*

Otro tipo de métricas empleadas en la detección de comunidades evalúan el nivel de agrupamiento de los nodos, si los nodos están estrechamente relacionados. El coeficiente de agrupamiento está relacionado con el número de triángulos.

Los **triángulos** en un grafo son los grupos conformados por tres nodos interconectados de manera total entre ellos, que conforman una triada [48].

El **coeficiente de agrupamiento o *clustering*** permite determinar el nivel de presencia de triángulos en una red. Este coeficiente se puede definir de dos formas, de manera global o de manera local [9].

**El coeficiente de agrupamiento global** también es conocido como **coeficiente de transitividad.** De manera formal, se define como [48]:

$$C_g = \frac{(\text{\# de triángulos en la red}) \times 3}{\text{\# de caminos de longitud 2 en la red}}$$

Para encontrar el número de triángulos en la red ($N_\Delta$) y el número de caminos de longitud 2 ($N_3$) se pueden utilizar las siguientes expresiones [9]:

$$N_\Delta = \sum_{k<j<i} a_{ij} a_{ik} a_{jk}$$

$$N_3 = \sum_{k<j<i} a_{ij} a_{ik} + a_{ji} a_{jk} + a_{ki} a_{kj}$$

Donde $a_{ij}$ son los elementos de la matriz de adyacencia $A$.

De manera similar, el **coeficiente de agrupamiento local** se calcula sobre un nodo en particular. De manera formal, se define como [48]:

$$C_i = \frac{\text{\# de triángulos conectados a } i}{\text{\# de caminos de longitud 2 conectando a } i}$$

También se puede encontrar este coeficiente en formulaciones mas amigables como [64]:

$$C_i = \frac{2m_i}{k_i(k_i - 1)}$$

Donde $m_i$ es el número de conexiones entre $k_i$ vecinos del nodo $i$.

También hay otra formulación para calcular el coeficiente de agrupamiento global a partir de la medida local [48]:

$$C_g = \frac{1}{n}\sum_{i \in V} C_i$$

### 2.4.5 Modularidad

El índice de **modularidad** fue propuesto por Girvan y Newman [52], para medir la presencia de *clusters* en una red no aleatoria. Se fundamenta en determinar las densidades dentro y fuera de cada *cluster* y comparándolas con la densidad general que tendrían si la red fuera construida de manera aleatoria [51].

De manera formal se define como:

$$Q = \frac{1}{2m}\sum_{ij}\left(a_{ij} - \frac{k_i k_j}{2m}\right)\delta(c_i, c_j)$$

Este $Q$ es llamado modularidad, $m$ es el número de aristas, $k_i$ es el grado del nodo $i$, $a_{ij}$ es el elemento de la matriz de adyacencia de los nodos $i$ y $j$, $\delta(m,n)$ es la función delta de Kronecker [2] y $c_i$ es el *cluster i*. [51]

Otra forma en la cual se presenta el índice de modularidad es la siguiente:

$$Q = \sum_r (e_{rr} - a_r^2)$$

Donde $e_{rr}$ son el conjunto de aristas que se conectan entre sí en un *cluster* denominado $r$ y $a_r$ representa al conjunto de nodos que forman parte del *cluster r* [51].

---

[2] La función delta de Kronecker toma el valor de 1 si las dos variables son iguales y 0 si las dos variables son diferentes. También puede presentar la notación $\delta_{ij}$ en la bibliografía.

## 2.5 Resumen

En este capítulo se estudia las características básicas para la comprensión del problema de la detección de comunidades en redes; se inicia con una revisión de los conceptos de grafo, adyacencia, peso y grado. También es importante la comprensión de las formas de representar este tipo de estructuras, de forma que sean más fáciles para su tratamiento y solución; también se revisó las estructuras básicas para la detección de una comunidad como lo son los subgrafos y entre estos los *cliques*.

Por último se realiza una revisión de las principales medidas implementadas para el estudio y detección de las comunidades, varias de estas son utilizadas por varios métodos, los cuales se revisaran en el capitulo siguiente.

Entre las medidas de detección se resaltan tres, la centralidad de un *cluster*, la cual puede ser evaluada desde varios enfoques como el grado del nodo, cercanía o intermediación; el coeficiente de agrupamiento de los nodos de manera global y local; y por último el índice de modularidad.

# 3. Métodos de detección de comunidades

El análisis del problema de agrupamiento o *clustering* es el proceso de dividir un conjunto de datos en subconjuntos. Cada subconjunto se denomina grupo o *cluster*, de tal forma que los objetos de un grupo son similares entre sí, pero sus datos se diferencian de otros *clusters*. El conjunto de *clusters* resultantes de este proceso se denomina agrupamiento. Dependiendo del método de agrupamiento o *clustering* aplicado, se pueden generar diferentes configuraciones de *clusters* para un mismo conjunto de datos [29].

Existen diferentes métodos de detección de comunidades dependiendo del tipo de definición con la cual se defina la comunidad (local, global y basada en similitud). Los métodos también se pueden diferenciar por el método de solución del problema, los cuales pueden ser enfoques jerárquicos, modulares o espectrales. Además, se puede considerar otra clasificación en función de si permiten detectar comunidades sin solapamiento o con solapamiento.

- **Comunidades sin solapamiento:** Este tipo comunidades se caracteriza por la generación de una cantidad determinada de comunidades, en las cuales cada individuo sólo pueden formar parte de una comunidad. Este tipo de comunidades se caracterizan por la presencia explícita de una jerarquía.

- **Comunidades con solapamiento:** En este tipo de comunidades los individuos pueden formar parte de más de una comunidad. Por ejemplo, las redes sociales un individuo puede formar parte de la comunidad "familia" y a su vez hacer parte de la comunidad "amigos".

Está clasificación no es totalmente excluyente; es decir, se pueden presentar métodos bajo un enfoque modular que considere solapamientos o un método jerárquico sin solapamientos.

En este capítulo, se expondrán diferentes métodos para la detección de comunidades desde los conceptos iniciales del proceso de agrupamiento, así como el enfoque utilizado por cada método, resaltando sus características principales. Por último, se resaltarán aquellas características que permitan hacer una comparación entre cada uno de los métodos empleados.

## 3.1 Métodos jerárquicos

Los métodos de agrupamiento jerárquicos consisten en establecer una jerarquía entre los distintos agrupamientos con diferentes niveles. Por lo general, este tipo de agrupamientos se visualizan en una estructura de árbol llamada dendrograma en donde las hojas representan los nodos individuales de la red y la raíz representa el conjunto total de nodos, construyendo un nivel de jerarquía entre los nodos individuales a grupos más pequeños, que luego se irán uniendo entre sí hasta formar un conjunto global.

En la **Figura 3-1** se puede observar el esquema de un dendrograma. Los círculos inferiores representan los nodos individuales de la red. A medida que se construye el modelo jerárquico, en el dendrograma se juntan los nodos, hasta llegar a un punto en el que se cumple algún criterio de parada, el cual se indica mediante la línea punteada horizontal. En este ejemplo, la red quedaría conformada por 4 agrupamientos [52].

**Figura 3-1:** Dendrograma [52]

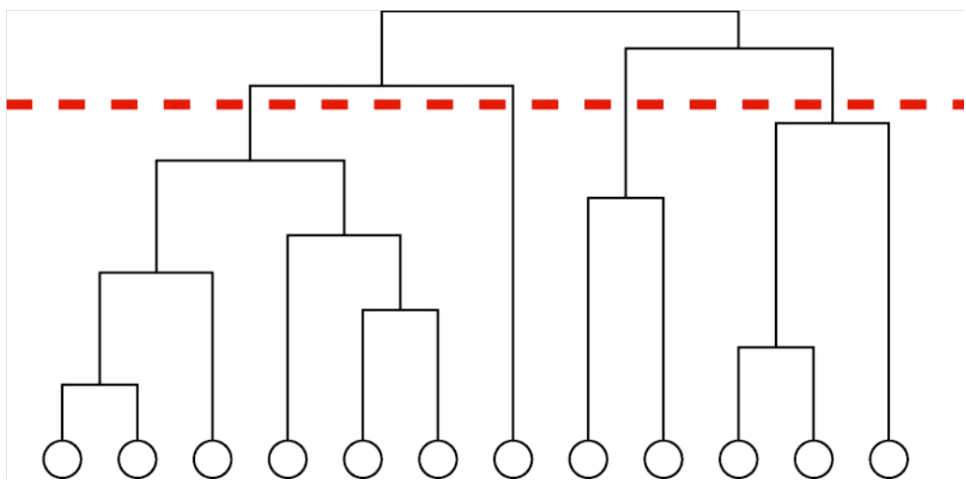

El objetivo fundamental de cualquier método jerárquico es definir la medida de similitud entre los nodos. Después, se establece una medida para cada uno de los nodos con el resto de la red, independientemente si se encuentran interconectados directamente o no. Por último, se genera una matriz $X$ de $n \times n$ (matriz de similitud) [19].

Comúnmente, se distinguen dos tipos de métodos jerárquicos para el agrupamiento [19]:

1. **Métodos divisivos:** Los grupos se conforman por medio de la eliminación de aristas con baja interacción entre los nodos.

2. **Métodos aglomerativos:** Los grupos de nodos se fusionan de manera iterativa, agrupando conjuntos de nodos con un alto grado de interacción.

A continuación, se expondrán algunos de los algoritmos más representativos que siguen este enfoque.

### 3.1.1 Métodos divisivos

Los métodos divisivos son métodos jerárquicos para la detección de comunidades. También conocidos como métodos *top-down*. Parten de considerar la red entera como un *cluster*, denominado *cluster* raíz. Luego el *cluster* raíz se divide en *clusters* más pequeños, los cuales se dividirán recursivamente hasta obtener cada nodo de la red como un *cluster* aislado [39].

#### 3.1.1.1  Método de Newman y Girvan

Uno de los algoritmos más populares es el introducido en el trabajo de Newman y Girvan que utilizan la medida de intermediación o *betweenness*. De manera general, esta medida consiste en identificar las aristas que cumplen la condición de puente entre cualquier par de nodos logrando la menor cantidad de saltos entre ellos [25]. Es de destacar que es muy importante este método porque marcó el inicio del estudio de detección de comunidades en el área de la Física [19].

De manera general, los pasos del algoritmo son [52]:

1. Calcular el índice de *betweenness* de todas las aristas de la red.

2. Eliminar las aristas con mayor índice de *betweenness* (en caso de empate, se escoge una de manera aleatoria).

3. Recalcular el índice de *betweenness* en toda la red.

4. Retornar al paso 2.

De una manera simplificada, su objetivo es encontrar aquellas aristas por las cuales existe un mayor "tráfico" o nivel de conexión entre nodos de la red que forman parte de diferentes comunidades. Cada vez que se encuentra una arista con altos valores de índice de *betweenness* esta se elimina y luego se recalcula el índice para todas las aristas de la red. Este proceso continuará de manera iterativa hasta que sobre la red se desarrollen subgrafos que se encuentren aislados del resto de la red. Este proceso se repite recursivamente sobre cada subgrafo hasta la construcción del dendrograma completo.

**Figura 3-2:** Identificación de aristas con alto grado de *betweenness* [51].

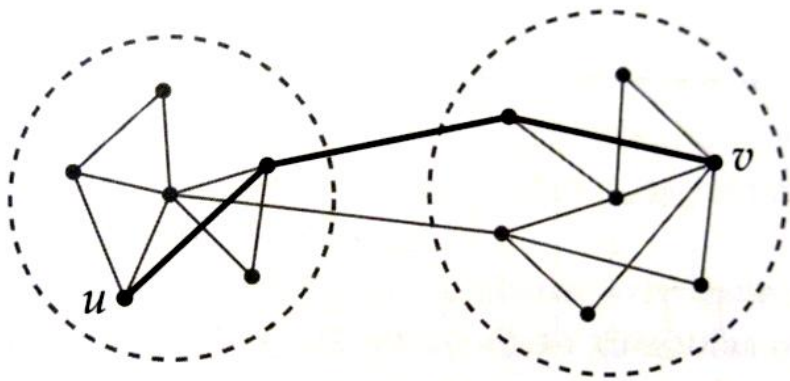

Uno de los grandes inconvenientes de este tipo de algoritmos es su complejidad computacional, la cual puede variar dependiendo del método empleado para evaluar la medida de *betweenness* [52]. Por lo general, esta complejidad está en el orden de $O(mn(m+n))$ dado un grafo $G$ con $m$ aristas y $n$ nodos, si bien puede llegar a $O(n^3)$ para redes donde $m \propto n$, por lo que este método no se aconseja en uso de redes con una alta cantidad de nodos y aristas [51]. A pesar de estos inconvenientes, cabe resaltar los buenos resultados que proporcional el algoritmo [52].

### 3.1.1.2 Método de Radicchi

El algoritmo de Radicchi [61] consiste en detectar aristas dentro de los agrupamientos donde se forman bucles. Como las comunidades son conjuntos de nodos altamente interconectados, es de esperar que aparezcan en ellas bucles de aristas y nodos. Por el contrario, las aristas que conectan nodos entre agrupaciones es poco probable que estén involucradas en bucles [49].

Basándose en estas idea, el algoritmo de Radicchi propone un **coeficiente de agrupamiento de aristas** en el que los valores pequeños permiten identificar qué aristas forman parte de un mismo agrupamiento [19].

Este coeficiente de agrupamiento de aristas se basa en el principio del coeficiente de agrupamiento de nodos propuesto por Watts y Strogatz en 1998, el cual consiste en la relación entre el número de triángulos que incluyen un nodo y el número de posibles triángulos que se pueden formar [19].

**Figura 3-3:** Método de Radicchi [51].

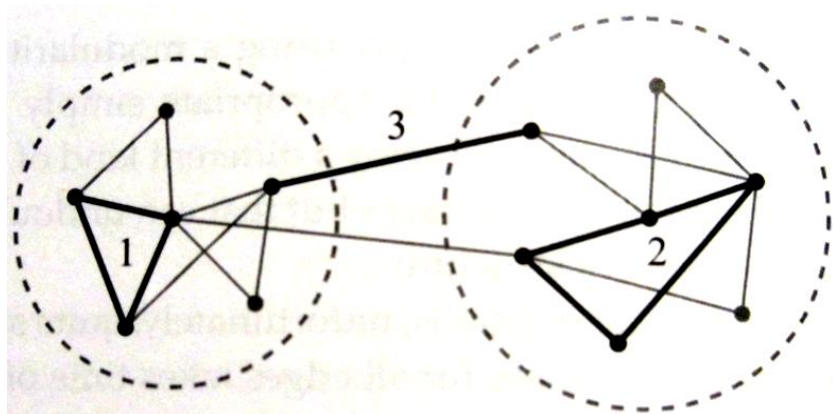

En la **Figura 3-3** se puede observar el principio del coeficiente de agrupación de manera grafica. El grado de atenuación permite identificar el valor del coeficiente de agrupamiento, cuanto más oscuro su coeficiente es mayor [19].

De manera formal, el coeficiente de agrupamiento se define como [49]:

$$C_{ij} = \frac{z_{ij} + 1}{\min(k_i - 1, k_j - 1)}$$

Donde $z_{ij}$ es el número de triángulos a los que los nodos $i$ y $j$ pertenecen y $k_i$ corresponde al grado del nodo $i$. Esta medida se basa en el hecho de que los bordes que conectan las comunidades tienden a exhibir un pequeño valor de éste coeficiente. El valor $+1$ presente en $z_{ij} + 1$ se utiliza para penalizar el caso en el que se presenten varias aristas con cero triángulos [49].

El coeficiente de agrupamiento tiene valores prácticamente inversos a los obtenidos por el algoritmo de Girvan y Newman: cuando se obtienen valores de *betweenness* altos, lo más probable es que se obtengan bajos valores de agrupamiento y viceversa [49].

Una ventaja de este método es su velocidad, su complejidad algorítmica es de orden $O(n^2)$. pero el algoritmo tiende a fallar si el coeficiente promedio de agrupamiento de la red es pequeño, entonces el coeficiente será pequeño para todas las aristas [51].

Este método tiene la ventaja de que funciona en redes con gran número de triángulos *k-cliques* en una primera instancia, lo que lo hace idóneo para implementarlo sobre redes sociales. Caso contrario ocurre con redes tecnológicas o biológicas, donde hay una menor cantidad de estas estructuras, por lo que es más difícil identificar las aristas entre grupos [51].

Determinar la centralidad es importante para el análisis de las redes en las que se propaga información de un nodo al resto de la red. Por lo general, se utilizan métodos heurísticos, grados de centralidad, algoritmos *greedy* o *betweenness*, entre otros. Trabajos propuestos por De Meo et al. profundizan en el concepto de *k-path* con la cual se calcula la importancia de las aristas en una red [12]. Las medidas tradicionales de centralidad pueden tener deficiencias al analizar redes con estructuras de comunidad en su interior, por eso Zhang et al. exponen una nueva forma de medir esta centralidad, por medio de la predicción de caminos de transferencia de información entre los nodos y el uso del algoritmo de agrupamiento *k-medoids* [90].

También hay trabajos que estudian una nueva forma de definición de la centralidad basados en el algoritmo de Girvan – Newman y la eliminación de aristas de manera iterativa. Se han propuestos varios métodos para calcular la centralidad en este caso, teniendo en cuenta topología, caminos y rutas [78].

Los métodos de clustering jerárquico se han basado principalmente en la combinación de diferentes resultados de diferentes *clusters* a partir de un único conjunto de datos. Hossain *et al.* proponen un método para consolidar los resultados del agrupamiento de varios conjuntos de datos, permitiendo complementar información que esté representada sobre varias redes en un único resultado [30].

### 3.1.2 Métodos aglomerativos

Los métodos aglomerativos se construyen de manera inversa a los métodos divisivos, también denominados *botton-up,* considerando cada nodo como un *cluster* y fusionan los pares de más similares (más cercanos) conformando *clusters* nuevos, este proceso se realiza iterativamente hasta conformar un único *cluster* [39].

El método de determinar las distancias entre los *clusters* es lo que diferencia los métodos aglomerativos. Los más representativos son:

- Conexión simple (*simple-link*).
- Conexión completa (*complete-link*).
- Conexión media (*average-link*).

De manera general, el algoritmo aglomerativo de *clustering* jerárquico consiste en [92]:

1. Iniciar todos los nodos $x_1, x_2, ..., x_n$ como *clusters* $C_1, C_2, ..., C_n$.
2. Juntar los clusters más cercanos ($C_i, C_j$).
3. Repetir el paso 2 hasta que sólo quede un *cluster* o se cumpla algún criterio de parada.

Uno de los problemas de los métodos aglomerativos es la tendencia a centrarse en los nodos con alto grado de conectividad, dejando un poco de lado el análisis de las zonas de la red más aisladas (ver **Figura 3-4**) [52].

**Figura 3-4:** Red analizada por un método aglomerativo [52].

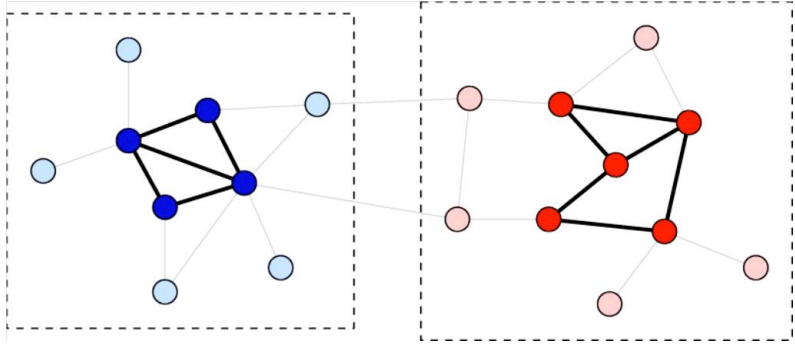

#### 3.1.2.1 Método de conexión simple (*simple-link*)

El método de conexión simple (*simple-link*) [18] para el *clustering* jerárquico consiste en encontrar la pareja de nodos más cercana (los dos nodos mas similares entre sí que pertenezcan a diferentes *clusters* y fusionar dichos *clusters*) [39].

De manera formal, la distancia mínima entre dos *clusters* es [92]:

$$d(C_i, C_j) = \min d(x, x'), x \in C_i, x' \in C_j$$

Uno de sus primeros trabajos se desarrolló en el algoritmo SLINK [73]. La complejidad computacional de este método es de $O(n^2)$, donde $n$ es el número de nodos.

Uno de los problemas de este método se conoce como el problema del encadenamiento. Este problema consiste en que se conforman grupos de forma elíptica, tal como se puede ver en la **Figura 3-5**.

**Figura 3-5:** Red analizada por un método aglomerativo simple [39].

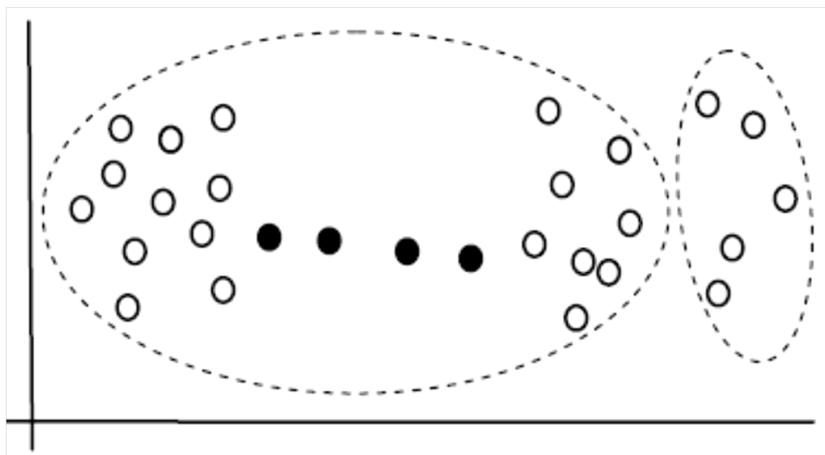

### 3.1.2.2 Método de conexión completa (*complete-link*)

El método de conexión completa (*complete-link*) para el *clustering* jerárquico es muy similar a método *simple link*, con la diferencia que su objetivo no es encontrar al nodo más cercano, sino encontrar conjuntos tipo *clique*. El método fusiona los dos *cluster* cuyos nodos más alejados tienen la distancia más pequeña [39].

De manera formal, la distancia máxima entre dos *clusters* es [92]:

$$d(C_i, C_j) = \max d(x, x'), x \in C_i, x' \in C_j$$

Uno de sus primeros trabajos se desarrolló en el algoritmo CLINK [13]. La complejidad computacional de este método es del orden de $O(n^2 \log n)$ donde $n$ es el número de nodos de la red.

Este método tiende a formar grupos más compactos y con diámetros similares, (ver **Figura 3-6**). También es más robusto frente el problema del encadenamiento, pero puede tener problemas con valores atípicos.

**Figura 3-6:** Red analizada por un método aglomerativo completo [39].

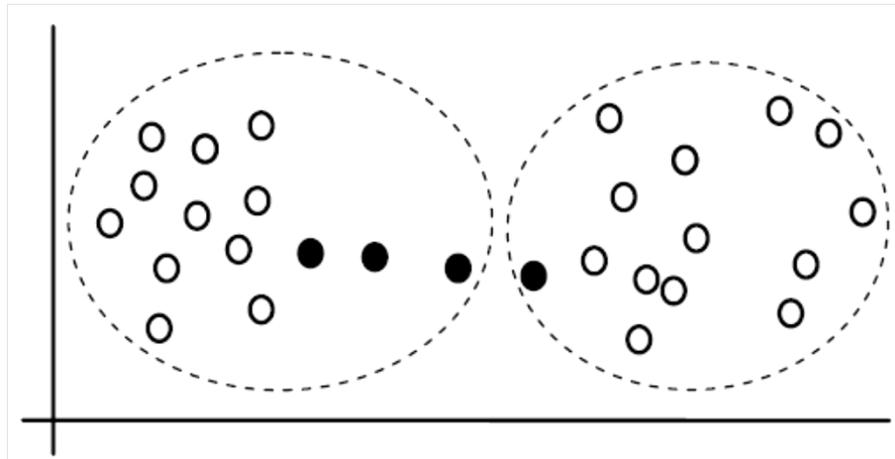

### 3.1.2.3 Método de conexión media (*average-link*)

El método de conexión media (*average-link*), también conocido como UMPGA [74], para el *clustering* jerárquico tiene en consideración el caso de los valores atípicos del método de conexión completa y la propiedad de encadenamiento de método simple. La distancia entre dos grupos se calcula como la distancia media de todas las distancias por pares entre los nodos en dos grupos [39].

De manera formal, la distancia media entre dos *clusters* es [92]:

$$d(C_i, C_j) = \frac{\sum x \in C_i, x' \in C_j d(x, x')}{|C_i| \cdot |C_j|}$$

La complejidad computacional de este método es del orden de $O(n^2 \log n)$, donde $n$ es el número de nodos de la red.

Aparte de las medidas de conexión analizadas anteriormente como los métodos de conexión simple, máxima y media; se emplean otros métodos basados en medidas geométricas o el cálculo de error cuadrático medio como los métodos de centroide, mediada y Ward [39].

Como se describió anteriormente, los algoritmos basados en Girvan-Newman presentan una alta complejidad computacional por las características del agrupamiento jerárquico. El gran volumen de datos representa una problemática constante. Li *et al.* proponen una modificación de la definición del coeficiente de agrupamiento de las aristas, desarrollando un método de agrupamiento rápido (FAG-EC) [44].

Uno de las grandes aplicaciones de este tipo de métodos es el análisis de datos procedentes de la biología. Wang *et al.* se centran en el análisis de las interacciones proteína – proteína, pero la implementación de técnicas jerárquicas no logra obtener buenos resultados dado el alto grado de ruido presente y los falsos positivos. Para afrontar el problema, proponen un método teniendo en cuenta medidas locales para las aristas en el empleo de un clustering jerárquico (HC-PIN), tanto para redes ponderadas como no ponderadas [31].

Para el análisis de redes sociales, Zhao *et al.* proponen un nuevo método para descubrir comunidades sobre redes sociales basado en el método jerárquico y complementado con la construcción de grafos ponderados, construcción de arboles más pequeños y consideración de *cliques* [91]. Algo similar proponen Shen *et al.* con su algoritmo EAGLE [71].

Una de las debilidades de los métodos de detección de comunidades, es análisis de redes con ruido o datos incompletos, lo que se representa en falta de aristas y/o nodos. Lin, et al. por medio de métodos jerárquicos proponen una aproximación de los datos faltantes en la red utilizando distancias locales, para su posterior análisis [38].

## 3.2 Métodos modulares

Los métodos modulares parten del criterio de modularidad establecido anteriormente por Girvan y Newman en su método, pero cada vez cuenta con más aceptación el término de "modularidad" como cualquier medida adecuada para determinar y encontrar agrupamientos. Por esta razón, autores como Fortunato han clasificado las técnicas basadas en esta medida como un grupo aparte, cuyo objetivo principal es encontrar agrupamientos que permitan optimizar el valor de modularidad y sean resuelvan el problema de optimización restante en tiempos razonables, dada su alta complejidad [19].

### 3.2.1 Método *Fast Greedy*

El algoritmo se basa en el algoritmo de Girvan y Newman. En su ejecución, se produce la división de los nodos en comunidades, independientemente si estas divisiones son naturales o no. Para determinar la calidad de las divisiones, se utiliza el concepto de modularidad $Q$, el cual se define de la siguiente manera [50]:

$$Q = \sum_i \left(e_{ii} - a_i^2\right)$$

donde $i, j$ son comunidades de la red, $e_{ii}$ son las aristas que hay en la comunidad $i$, y $a_i = \sum_j e_{ij}$.

Si al momento de realizar las divisiones en la red, el valor resultante es menor que la cantidad esperada al azar, esta modularidad es $Q = 0$. Valores distintos de 0 indican desviaciones de la aleatoriedad y con valores superiores a 0.3 se aprecia una estructura de comunidad significativa [50].

La complejidad computacional de este método es del orden de $O((m + n)n)$, donde $n$ es el número de nodos y $m$ es el número de aristas [50].

### 3.2.2 Otros métodos

Los métodos modulares se basan principalmente en el método anterior. A partir de él, se han propuesto mejores que se centran en el desarrollo de técnicas rápidas, ideales para redes de gran tamaño que calculan valores aproximados a $Q$; técnicas más precisas, pero con alto costo computacional y técnicas intermedias que intentan alcanzar un compromiso entre velocidad y precisión.

Este tipo de métodos presenta dos deficiencias para la solución de problemas de detección de comunidades [19]:

- Cuando los *clusters* no tienen un tamaño homogéneo.
- La presencia de *clusters* no comunes, (se pueden generar comunidades que so sean coherentes al problema analizado).

Sobre el concepto de modularidad, se han desarrollado varios métodos, como el propuesto por Clauset *et al.* que implementan el mismo principio de los métodos *greedy*, pero aplicando estructuras de datos más complejas y permitiendo una ejecución con una complejidad de $O(m \log^2 n)$ [7].

Ovelgönne y Geyer-Shulz presentan un método llamado *randomized greedy (RG)* [56]. Su objetivo es obtener una buena modularidad en redes de millones de nodos y con bajo costo computacional. Para lograr este equilibrio, se fundamenta en la implementación de técnicas heurísticas [22]. Otro método basado en la modularidad es el FPMQA el cual parte de una detección de *clusters* iniciales y la fusión de los mismos basada en medidas de modularidad [6].

## 3.3 Métodos particionales

Los métodos particionales predisponen la cantidad de grupos o *clusters* que se desean encontrar, $k$. Los datos son modelados en un espacio métrico donde cada nodo es representado por medio de un punto en el espacio y una medida de distancia es definida entre dos puntos. Está distancia representa la similitud entre los puntos [19].

El objetivo es separar los puntos en $k$ grupos maximizando o minimizando una función de costo basándose en las distancias que existen entre los puntos o los centroides de los grupos, de manera iterativa. Uno de los algoritmos más representativos de este tipo de técnicas es el algoritmo de las $k$-medias (*k-means*) [19].

También se incluyen en este tipo los métodos que tratan de dividir una red en conjuntos de nodos, donde la cantidad de aristas entre los grupos sea la mínima [19].

**Figura 3-7:** Red particionada por método grafico [19].

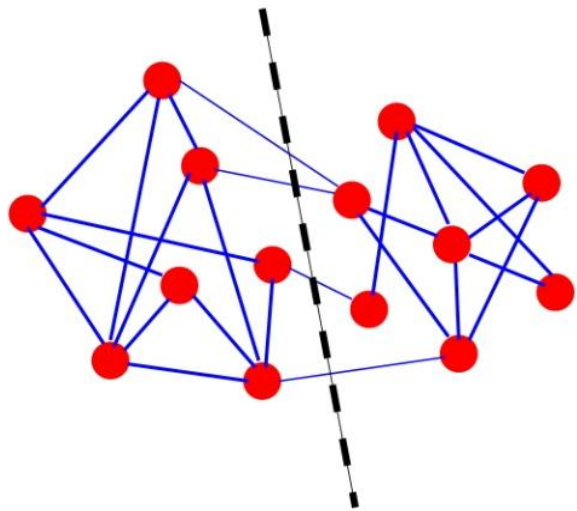

En la **Figura 3-7** se puede ver el ejemplo de una partición de tamaño 2: conseguir dos grupos de igual tamaño con el menor número de aristas entre ellos [19].

El particionamiento de grafos es un tema fundamental de investigación en la computación paralela y en el diseño de circuitos. Otra de las áreas de aplicación de este tipo de métodos es la resolución de problemas de ecuaciones diferenciales parciales y sistemas de ecuaciones lineales dispersos [19].

Por lo general, los problemas de particionamiento de grafos son NP-duros. Existen variaciones para mejorar su rendimiento, aunque las soluciones que proporciona no son del todo precisas. Muchos de estos algoritmos emplean técnicas de bisección iterativa [19].

### 3.3.1 *k*-medias

El algoritmo de k-medias fue desarrollado por McQueen en 1967 [40] y, tal como se mencionaba anteriormente, es uno de los más utilizados por su simplicidad [19].

El algoritmo de k-medias parte del número de grupos a encontrar ($k$). En cada grupo se tiene un centroide, el cual es el centro geométrico del *cluster*. Por lo general, este centroide representa al *cluster* en su totalidad, y no es más que la media aritmética o media ponderada de los integrantes del *cluster* [39].

El algoritmo de las k-medias consiste en los siguientes pasos [39].

1. Escoger aleatoriamente $k$ puntos iniciales como centroides iniciales.

2. Calcular la distancia entre cada centroide y cada dato. Cada punto es asignado a su centroide más cercano y el conjunto de puntos asociados a un centroide conforman un *cluster*.

3. Recalcular los centroides a partir del conjunto de puntos asignados a cada *cluster*.

4. Repetir pasos 2 y 3 hasta que se cumpla un criterio de parada.

**Figura 3-8:** Conjunto inicial y final de datos, método k-medias [39]

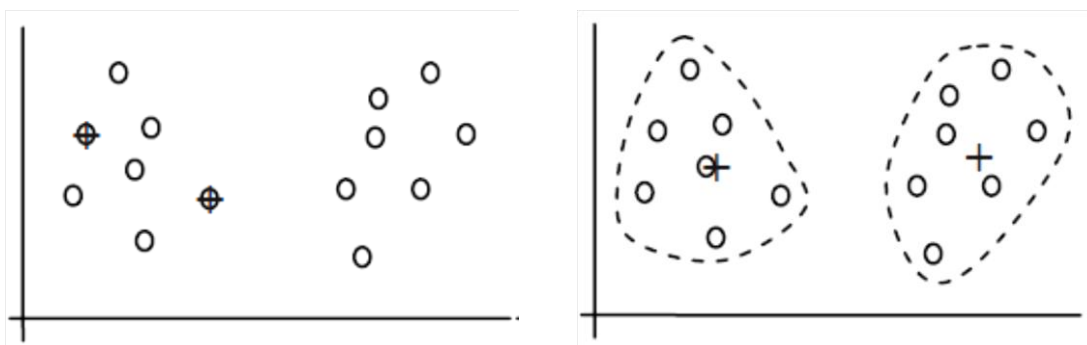

(A)  Selección aleatoria de las k semillas.

(B) *Clusters* finales con sus centroides

En la **Figura 3-8** se puede apreciar el estado inicial del algoritmo para un pequeño conjunto de datos y los grupos encontrados al final del proceso. Para llegar al estado final fueron necesarias 3 iteraciones en las cuales el centroide se recalculaba en función de la asignación [39].

Por lo general, los criterios de parada que se suelen utilizar son los siguientes [39]:

- No hay (o mínima) reasignación de puntos de datos a diferentes grupos.

- No hay (o mínimo) cambio en los centroides.

- Disminución mínima en la suma de errores cuadráticos:

$$SSE = \sum_{j=1}^{k} \sum_{\mathbf{x} \in C_j} dist(\mathbf{x}, \mathbf{m}_j)^2$$

donde $k$ es el numero de de clusters requeridos, $C_j$ es el $j$ ésimo *cluster*, $\mathrm{m}_j$ es el centroide del *cluster* $C_j$ y $dist(\mathrm{x}, \mathrm{m}_j)$ es la distancia entre el punto x y el centroide $\mathrm{m}_j$ [39].

En un espacio euclídeo, la media de un *cluster* se calcula como [39]:

$$\mathrm{m}_j = \frac{1}{|C_j|} \sum_{\mathbf{x}_i \in C_j} \mathbf{x}_i$$

donde $|C_j|$ es el número de puntos en el *cluster* $C_j$.

La distancia de cada punto $\mathbf{x}_i$ al centroide $\mathrm{m}_j$ se calcula por: [39]

$$dist(\mathrm{x}_i, \mathrm{m}_j) = \|\mathrm{x}_i, \mathrm{m}_j\| = \sqrt{(x_{i1} - m_{j1})^2 + (x_{i2} - m_{j2})^2 + \cdots + (x_{ir} - m_{jr})^2}$$

Uno de los inconvenientes de este método es que pueden aparecer *clusters* vacíos durante el proceso, cuando ningún dato fue asignado al *cluster*, lo que puede distorsionar los resultados del algoritmo [39].

Otro inconveniente es la indicación inicial del número de *clusters*, lo que limita las capacidades del algoritmo, ya que en varios problemas no es posible conocer con anterioridad el número de *clusters* que se desean encontrar. [19]

La complejidad de este método es relativamente baja, del orden de $O(tkn)$ donde $n$ es el número de datos, $k$ es el número de *clusters* y $t$ es el número de iteraciones, para las cuales se suele establecer el límite de iteraciones [39].

### 3.3.2 Algoritmo de Kernighan – Lin

El método de Kernighan – Lin es uno de los primeros empleados para el particionamiento de grafos. Kernighan y Lin lo propusieron en 1970 [33] como un método para la optimización de circuitos. Su objetivo es dividir el circuito en módulos que se interconectaran con el menor número de caminos, lo que facilitaba su ensamblaje y funcionalidad con respecto a la distribución de los componentes y el diseño general [19].

Sea $G = (V, E)$ un grafo con $2n$ nodos numerados de 1 hasta $2n$. Cada arista $(i, j)$ tiene un costo asociado $c_{ij}$. $C = (c_{ij})$ es la matriz de costo de $G$, donde $c_{ij}$ es el costo de ir del nodo $i$ al nodo $j$ si existe camino, en caso contrario, su costo es cero (0) [24].

Se realiza una partición del grafo $G$ en dos subgrupos $A, B$ de igual tamaño, cuyo objetivo es encontrar el menor número de caminos entre los dos subconjuntos. Para esto se necesita encontrar el mínimo costo de separación de las aristas intercambiando pares de nodos entre los subgrupos. La realización de estos intercambios hace que aumente o disminuya el número de las aristas que conectan los dos subgrupos, lo que se representa mediante una función de ganancia $g$ [24].

Las aristas que conectan un nodo del grupo $A$ al grupo $B$ y viceversa son llamadas *outedges* (aristas externas) y aquéllas que conectan los nodos de un mismo grupo son llamadas *inedges* (aristas internas). Sea $T$ el costo total de todas las *outedges*, el objetivo del algoritmo es reducir el valor de $T$ en cada iteración con el intercambio de los nodos de cada subgrupo [24].

El costo externo de un nodo $a$ se define como [24]:

$$E_a = \sum_{x \in B} c_{ax}$$

De manera reciproca, se define el costo interno [24]:

$$I_a = \sum_{x \in A} c_{ax}$$

La ganancia $g$ se obtiene a partir del parámetro $D$. Este parámetro se obtiene a partir de la diferencia entre el costo externo de un nodo y su costo interno [24].

$$D_a = E_a - I_a$$

Esta ganancia es el valor del intercambio entre $a$ y $b$ [24].

$$g(a,b) = D_a + D_b - 2c_{ab}$$

Al iniciar el método, todos los nodos son libres de intercambiarse entre los grupos. A medida que va avanzando el método, se bloquean los nodos que son intercambiados y se continúa con aquellos que están libres. Termina el algoritmo cuanto todos los nodos quedan bloqueados. Al terminar el intercambio se desbloquean los nodos y se repite el proceso nuevamente [24].

El algoritmo se resume a continuación [24]:

1. Realizar la partición del grafo $G$ en dos subgrafos con la misma cantidad de nodos $(A, B)$.

2. Seleccionar los nodos $a_i$ y $b_i$, que pertenecen a los subgrupos $A$ y $B$ respectivamente, de tal forma que se maximice la ganancia cuando $a_i$ es intercambiado por $b_i$. Cuando no haya más nodos que comprobar se continúa con el paso 3.

    a. Si hay una mejora en el intercambio entre $a_i$ y $b_i$, realizar el intercambio y bloquear los nodos para impedir desplazamientos retornando al paso 2.

    b. Si no hay mejora, incrementar el contador i y retornar al paso 2.

3. Volver al paso 1 y repetir el procedimiento hasta que la ganancia sea mínima.

El algoritmo de Kernighan – Lin es lento, con una complejidad computacional promedio de $O(mn^2)$ donde $n$ es el número de nodos y $m$ el número de aristas; esta puede aumentar incluso a un orden de $O(n^4)$ en el caso de redes densas [51].

Además, este método tiende a decantarse por los resultados locales, por lo que no es tan recomendable utilizarlo como método inicial bajo un particionamiento aleatorio. Se obtienen mejores resultados si se parte de un particionamiento supervisado o como método de depuración de otros métodos de agrupamiento [51].

Este método se utiliza bastante para determinar puntos de corte adecuados. En este caso, Barkoosarae *et al.* implementan el algoritmo de Kernighan – Lin junto con algoritmos heurísticos para dar solución al problema de hallar *clusters* en base a nodos móviles y así evitar el problema de cuellos de botella en las comunicaciones de la red móvil bajo la tecnología HMIPv6 [2]. Sahu *et al.* aplican el algoritmo LMAP, basado en algoritmos heurísticos y Kernighan – Lin, para el diseño más adecuado de sistemas NOC (*Network-on-chips*) basado en una topología de malla, con el objetivo de mejorar los tiempos de latencia entre los nodos de la red [69].

## 3.4 Métodos espectrales

Los métodos espectrales se proponen para dar solución al problema de la partición de grafos mediante el uso de vectores propios, de tal forma que las comunidades sean lo más disjuntas posibles [19].

El método consiste en [19]:

1. Transformación del conjunto de nodos, en un conjunto de puntos de un espacio métrico, cuyas coordenadas son elementos de vectores propios.

2. La agrupación del conjunto de puntos se agrupa por medio de técnicas como el método de las k-medias.

El objetivo es encontrar diferentes *clusters* con bajos grados de similitud entre ellos pero con altos grados de similitud internos. Existen varios métodos pero se basan en el mismo principio.

### 3.4.1 Conceptos preliminares

Si el grafo se partiera en dos conjuntos de nodos $A, B \subset V$ y $A \cap B = \emptyset$, entonces [81]:

$$w(A, B) = \sum_{i \in A, j \in B} w_{ij}$$

Hay dos formas para determinar el tamaño del subconjunto $A \subset V$ [81]:

$$|A| = \text{número de nodos en } A$$

$$\text{vol}(A) = \sum_{i \in A} d_i$$

Para determinar los parámetros clave para el *clustering* espectral, se debe tener en cuenta el desarrollo de la matriz laplaciana y el concepto de similitud, el cual se trata a profundidad en el artículo *A Tutorial on Spectral Clustering* de Ulrike von Luxburg [81].

### 3.4.2 Laplaciano de un grafo

Los laplacianos de un grafo son la principal herramienta del clustering espectral. De hecho, existe un campo de investigación sólo en el estudio de las matrices laplacianas [81].

De manera formal, la **matriz laplaciana no-normalizada** se define como [81].:

$$L = D - W$$

donde $W$ es la matriz de pesos y $D$ es la matriz diagonal de grados.

Hay dos tipos de **matrices laplacianas normalizadas,** las cuales se definen como [81].:

- Normalización por camino aleatorio:

$$L_{\text{rw}} = D^{-1}L = I - D^{-1}W$$

- Normalización simétrica:

$$L_{\text{sym}} = D^{-\frac{1}{2}}LD^{-\frac{1}{2}} = I - D^{-\frac{1}{2}}WD^{-\frac{1}{2}}$$

### 3.4.3 Enfoques de particionamiento

Varios algoritmos de agrupamiento se basan en problemas de particionamiento gráfico. Entre estos se encuentran los algoritmos espectrales de agrupación. Estos algoritmos se fundamentan en la solución del problema de realizar el corte de un grafo en dos grafos, que luego se generaliza a varios $n$-arios. El particionamiento $n$-ario consiste en eliminar las aristas de un grafo para encontrar $k$ subgrafos independientes [47].

A continuación se explicarán los fundamentos para particionar un grafo en dos:

#### 3.4.3.1 Corte mínimo *(MinCut)*

El objetivo del método de corte mínimo consiste en encontrar dos clusters $A$ y $B$, cuyas conexiones entre ellos sean del menor peso posible. La función objetivo se define como:

$$cut(C_1, \ldots, C_k) = \frac{1}{2} \sum_{i=1}^{k} W(C_i, \overline{C_i})$$

donde

$$W(C_1, C_2) = \sum A(i,j), i \in C_1, j \in C_2$$

$k$ es el número de subgrafos y $(C_1, \ldots, C_k)$ es el conjunto de posibles *clusters*.

Uno de los inconvenientes de esta medida es su aplicación sobre grafos desbalanceados, lo cual puede provocar *clusters* poco razonables [28].

#### 3.4.3.2 Mínimo radio de corte *(RadioCut)*

En el método de *RadioCut* [27], propuesto por Hagen y Kahng, se elimina el requisito de que los *clusters* tengan que ser balanceados. El tamaño del *clusters* es medido por el número de nodos [28]:

$$RadioCut(C_1, \ldots, C_k) = \frac{1}{2} \sum_{i=1}^{k} \frac{W(C_i, \overline{C_i})}{|C_i|} = \sum_{i=1}^{k} \frac{cut(C_i, \overline{C_i})}{|C_i|}$$

### 3.4.3.3 Corte mínimo normalizado *(NCut)*

En este caso el corte mínimo normalizado[72], propuesto por Shi y Malik, se propone que el tamaño del *cluster* $C$ se mida por los pesos de sus aristas vol($C$) [28]:

$$NCut(C_1, \ldots, C_k) = \frac{1}{2}\sum_{i=1}^{k} \frac{W(C_i, \overline{C_i})}{\text{vol}(C_i)} = \sum_{i=1}^{k} \frac{cut(C_i, \overline{C_i})}{\text{vol}(C_i)}$$

### 3.4.3.4 Corte Min-Max *(MinMaxCut)*

El corte min-max [14], propuesto por Ding *et al.* permite obtener resultados más compactos y balanceados que el *RadioCut* y *NCut*. En el método de corte Min-Max, las similitudes dentro de los *clusters* se maximizan de forma explícita [28]:

$$MinMaxCut(C_1, \ldots, C_k) = \frac{1}{2}\sum_{i=1}^{k} \frac{W(C_i, \overline{C_i})}{W(C_i, C_i)} = \sum_{i=1}^{k} \frac{cut(C_i, \overline{C_i})}{W(C_i, C_i)}$$

## 3.4.4 Algoritmo básico de particionamiento espectral

El concepto principal de los algoritmos espectrales es la consideración de un conjunto de $n$ datos $(x_1, \ldots, x_n)$, para los que se miden sus similitudes. La similitud entre parejas de datos, viene dada por una función de similitud que es simétrica y no negativa [81].

Entradas: matriz de similitud $S$, $k$ clusters a construir

- Construir el grafo de similitud.
- Calcular la matriz laplaciana.
    - Si es *clustering* no normalizado, por medio de $L$.
    - Si es *clustering* normalizado, por medio de $L_{\text{rw}}$.
- Calcular los primeros $k$ vectores propios $v_1, \ldots, v_k$ de la matriz laplaciana.
- Construir la matriz $V \in \mathbb{R}^{n \times k}$, donde los vectores propios son las columnas.
- Interpretar las filas de $V$ como un nuevo conjunto de puntos $A_i \in \mathbb{R}^k$.

- Agrupar los puntos $A_i$ con el algoritmo de las k-medias en $\mathbb{R}^k$

Salida: *Clusters* $A_1, \ldots, A_k$ con $A_i = \{j | y_j \in C_i\}$.

A partir de este algoritmo, y dependiendo si el grafo está normalizado o no, así como también de la función empleada para realizar el corte, se han propuesto varios algoritmos, algunos de ellos son:

- Algoritmo sin normalizar [15].
- Algoritmo NG, Jordan y Weiss [53].
- Algoritmo Shi y Malik [72].

La complejidad de este tipo de métodos es bastante alta, tanto en recursos de cómputo como en memoria utilizada dado el alto número de operaciones a realizar. De manera general, presenta una complejidad en tiempo de computo de $O(n^3)$ [19] y $O(n^2)$ en requerimientos de memoria [75].

### 3.4.5 Modificaciones

Como la mayoría de los algoritmos espectrales usan técnicas de agrupamiento como el algoritmo de las k-medias, lo que conduce a converger preferiblemente a un optimo local, Wang *et al.* proponen la combinación de algoritmos genéticos con los métodos espectrales, cambiando el método de las k-medias por un método k-medias genético que evita que converja rápidamente a un óptimo local [84].

Yu *et al.* indican la adecuada selección de la matriz de similitud, que al basarse en distancias euclídeas, puede obtener malos resultados cuando se presentan valores atípicos o datos con mucho ruido. Debido a esto, se buscan nuevas formas de plantear las distancias entre los puntos y la construcción de matrices de similitud, obteniendo como resultado un suavizado de los datos, lo que los hace más resistentes a estos fenómenos [89].

Otro campo de estudio de los algoritmos espectrales es determinar un parámetro adecuado de escala para determinar el grado de similitud entre los nodos y, a pesar del amplio estudio sobre estas áreas, Wu *et al.* consideran que se puede seguir estudiando debido a las fallas que presentan estos métodos sobre algunos escenarios como los expuestos anteriormente [86].

Una de las grandes deficiencias de los algoritmos espectrales es su escalabilidad. Al basarse en operaciones matriciales, existen investigaciones que permiten aprovechar las capacidades de la computación paralela para aplicar sobre este tipo de modelos [75]. Dados los avances de la actualidad en lo correspondiente a computación distribuida y paralela, muchas de estas técnicas se están optimizando para aprovechar las ventajas que ofrecen, este es el caso que proponen Wen-Yen *et al.* estudiando la paralelización del algoritmo en base al manejo de los vectores propios, lo que permite afrontar problemas de grandes dimensiones. [85]

Tabatabaei *et al.* proponen complementar el procedimiento del método espectral con algoritmos de agrupamiento jerárquico aglomerativo y refinamiento local. A este método se le conoce como GANC; presenta una menor complejidad computacional, lo que permite analizar mayores volúmenes de datos [19].

## 3.5 Detección de comunidades con solapamiento

La presencia de comunidades solapadas (*overlapping communities*) es un fenómeno común cuando se analizan redes sociales. Por ejemplo, en una red social, una persona puede pertenecer a varias comunidades como su familia, amigos, *hobbies*, trabajo; encontrar este tipo de comunidades con las técnicas tradicionales es algo para lo que no están preparadas [19].

En la **Figura 3-9**, se puede ver como un nodo pertenece a varias comunidades. Estas comunidades no son únicas, también forman parte de comunidades más grandes, lo que significa que éstas se solapan entre sí. Este fenómeno ha sido estudiado ampliamente por las ciencias humanas. En la actualidad, con el auge de las redes sociales en Internet, este tipo de problemas está empezando a ser estudiado por las ciencias de la computación [19].

**Figura 3-9:** Concepto de comunidades solapadas[57].

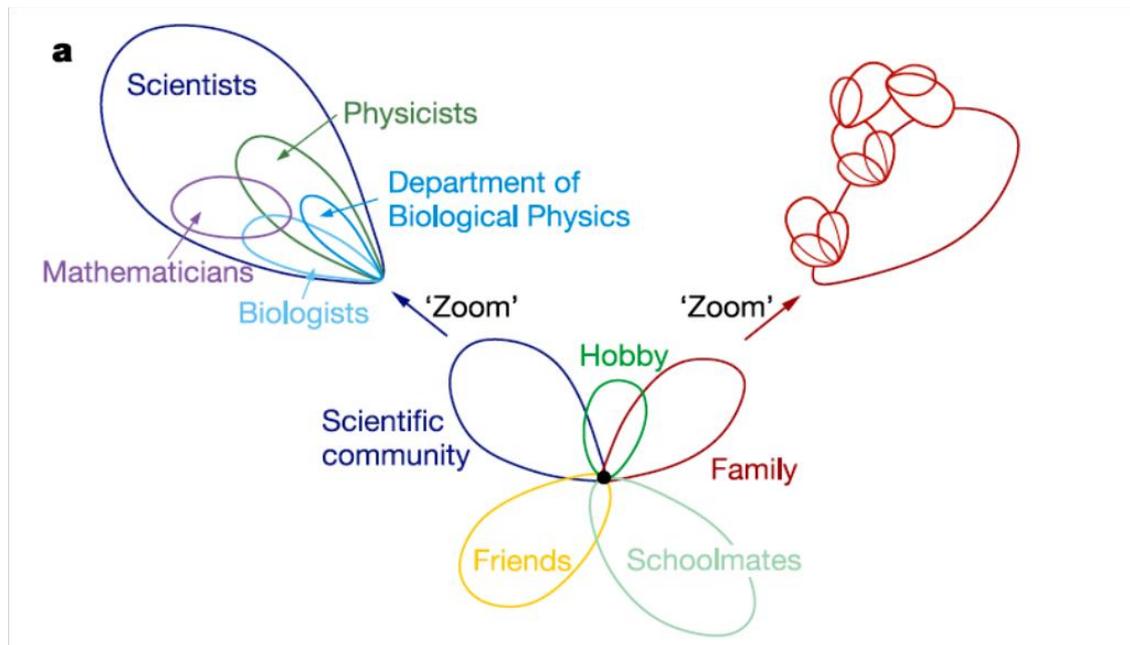

Las técnicas descritas en las secciones anteriores están enfocadas a la detección clásica de comunidades, en las que un nodo puede pertenecer exclusivamente a una comunidad. A continuación se expondrá algunas de las técnicas más representativas para la detección de comunidades solapadas [19].

### 3.5.1 Adyacencia de subconjuntos

Dos conjuntos se consideran adyacentes si se superponen entre sí con tanta fuerza como sea posible; es decir, si comparten $k-1$ nodos. Si se elimina un enlace de un *k-clique*, esto conduce a la formación de dos $(k-1)$-*cliques* adyacentes que comparten $(k-2)$ nodos [19].

La unión de estos conjuntos adyacentes conforma una cadena de *k-cliques*. Dos *cliques* se encuentran conectados si forman parte de la misma cadena. Finalmente, una comunidad la conforman el conjunto de *k-cliques* interconectados sobre una misma cadena [19].

En la **Figura 3-10** se puede ver un ejemplo de una comunidad solapada con $k = 4$. La comunidad inferior derecha (amarilla) se encuentra solapada con la comunidad superior derecha (azul) a través de un nodo. En cambio, se encuentra solapada con la comunidad inferior izquierda (verde) a través de tres nodos y un enlace entre dos de los nodos compartidos [57].

**Figura 3-10:** Ejemplo de subconjuntos solapados en una comunidad [57].

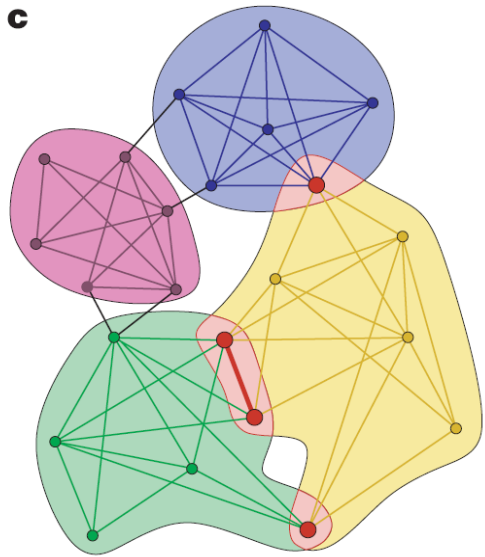

### 3.5.2 *Clique Percolation Method*

Una de las técnicas más populares es la propuesta por Palla *et al.* en 2005 [57] llamada método de percolación de *cliques* o, en inglés, *clique percotation method (CPM)*, el método se basa en el fenómeno de que las aristas en un grafo altamente conectado tiende a formar *cliques*. De manera opuesta, las aristas que conectan nodos de diferentes comunidades no tienden a formar este tipo de estructuras. Este concepto tiene una base similar a la propuesta para el método de agrupamiento de Radicchi [19].

Las comunidades detectadas usando este método basadas en la idea clave de formación de *cliques*, se construyen a partir de bloques adyacentes de un mismo tamaño $k$, que corresponden a *k-cliques.*

#### 3.5.2.1 Método

El algoritmo CPM consiste en encontrar los *k-cliques* adyacentes que permitan conformar una cadena. Cuando no sea posible extender la cadena de adyacencias, se da por formada la comunidad.[19]

Dentro de esta comunidad, es posible "rotar" o "pivotar" los *k-cliques* al largo de la cadena tan solo reemplazando un nodo del *clique* [19].

**Figura 3-11:** Proceso de formación de una comunidad por CPM [23].

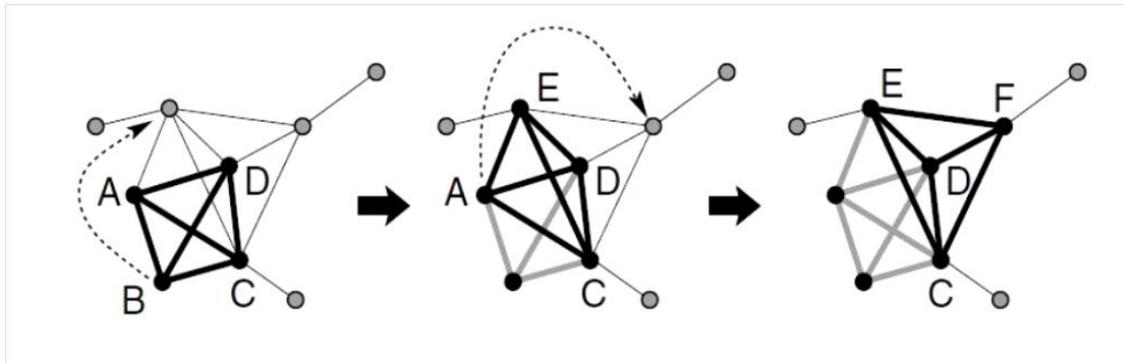

En la **Figura 3-11** se puede apreciar el proceso de formación de manera más didáctica. En el paso 1 se tiene un *clique* con un $k = 4$ formado por los nodos $A - B - C - D$. En el paso 2 se "rota" a un nodo adyacente cumpliendo con la definición de adyacencia, esta rotación da lugar a un *clique* formado por los nodos $A - E - D - C$; este proceso se repite hasta que no sea posible continuar con las "rotaciones". En el paso 3 se puede ver la última formación de nodos $E - D - C - F$. Después de terminar el proceso, se obtiene una cadena de *cliques* que da lugar a una comunidad formada por los nodos $A - B - C - D - E - F$ [23].

De manera general, el algoritmo CPM consiste en tres pasos [23]:

- Buscar todos los *cliques* de tamaño $k$ en todo el grafo.

- Construir cadenas de *cliques*.

- Todos los nodos que intervienen en los *cliques* de la cadena forman parte de la comunidad.

Para encontrar los *cliques* adyacentes, se emplea una matriz de adyacencia de *cliques*, cuyo tamaño corresponde al número de *cliques* presente en el grafo. A continuación, se representa el número de nodos compartidos por cada par de *cliques*. Por último, se eliminan de la matriz de adyacencia aquellas celdas cuyo valor sea menor o igual a ($k-1$). Con esta matriz, las comunidades solapadas se pueden determinar fácilmente [19].

Este algoritmo tiene una complejidad exponencial, determinada por el número de *cliques* presentes en la red y el tamaño de la red (nodos, aristas). El número de *cliques* es muy difícil de determinar con antelación, pero en varios casos donde las redes contaban con $10^5$ nodos se puede resolver el problema en tiempos razonables [19].

En la actualidad, hay una implementación práctica del método CPM llamada CFinder[3]. Basada en el trabajo de Palla *et al.* [57], se ha determinado que su complejidad computacional es del orden de $O(n_c^2)$, donde $n_c$ es el número máximo de *cliques* presentes en la red [66].

### 3.5.2.2 Consecuencias

La definición de comunidad a partir de los *k-cliques* genera algunas consecuencias. Una de ellas es excluir aquellos nodos con pocas conexiones, que a su vez los excluyen de las comunidades altamente conectadas y, como consecuencia, no se tienen en cuenta en el momento de conformar las comunidades [60].

Este método tiene que ser utilizados asumiendo que la red a analizar contiene un alto contenido de *cliques*. Este método tiende a fallar en redes con pocos *cliques* en su interior. En caso contrario, con comunidades altamente interconectadas y, por ende, con varios *cliques*, se puede presentar que sólo se encuentre una sola comunidad. Lo esencial es determinar un valor adecuado para $k$, lo que dependerá de las características del problema concreto [23].

El parámetro $k$ es elegido de acuerdo a las necesidades del usuario. La identificación de comunidades con valores pequeños de $k$ es adecuada para conformar comunidades amplias en cobertura. En cambio, si se eligen valores altos de $k$, estos son más apropiados para el estudio de comunidades muy densas [60].

---

[3] http://www.cfinder.org/

### 3.5.3 CPM Secuencial

Este método, propuesto por Kumpula *et al.* [35] funciona de manera opuesta al CPM. Se basa en la idea de buscar comunidades de *k-cliques* por medio de la inserción de aristas en un grafo vacío. Cada vez que se añade una nueva arista al grafo, se comprueba la formación de *k-cliques*. A la par de la detección de los *k*-cliques, se construye el dendrograma correspondiente. El algoritmo ha sido especialmente diseñado para el análisis de redes ponderadas. También es posible su uso en redes no ponderadas, donde la arista por analizar es seleccionada de manera aleatoria [19].

El método se compone principalmente de dos pasos:

- Detección de *k-cliques* por la inserción de un enlace.
- Seguimiento y fusión de k-comunidades mediante el procesamiento de los *k-cliques*.

Todo el proceso se repite por cada *k-clique* introducido. Una vez se han formado los *k-cliques*, las k-comunidades pueden ser identificadas en el grafo elaborado. Además, se puede hacer el seguimiento al proceso de formación de *cliques* y detenerlo en cualquier momento, lo que puede ser útil para redes muy densas.

La complejidad computacional de este método es prácticamente lineal, con respecto al número de *k-cliques*.

### 3.5.4 Otros métodos

Una de las grandes debilidades del método CPM es su rendimiento a medida que crece la cantidad de nodos y, sobre todo, con el crecimiento de interconexiones. Por esta razón varios investigadores hacen diversas propuestas para dar solución a esta problemática, como el método MOSES [41] basado en modelos estadísticos para la estructura de las comunidades.

Otro método propuesto por Shang *et al.* se enfoca en la detección de los núcleos de las comunidades por medio de la detección de *maximal cliques*. A continuación se fusionan los núcleos para conformar una sola comunidad [70]. La simplificación de las redes para analizar por medio del análisis de subredes permite evaluar grandes conjuntos de datos, del orden de $10^6$ nodos [46].

Wu *et al.* proponen un método de dos etapas para afrontar el problema en redes muy grandes. Su método encuentra comunidades no solapadas y luego busca los vínculos entre ellas [87].

### 3.5.5 Otros enfoques

El problema de detectar comunidades solapadas es muy importante, como se estudió anteriormente. Existen varios enfoques como el método basado en difuminación o "fuzzificación", que permite modelar interacciones difusas. Este método emplea una relación difusa para describir la interacción entre los nodos y la topología de la red para modelar el grado de pertenencia de la relación [77].

La investigación en la detección de comunidades no sólo se centra en el desarrollo de algoritmos que permitan detectar eficazmente estas comunidades. Otro enfoque, propuesto por De Meo *et al.* para mejorar el proceso de detección, consiste en establecer una etapa de preprocesamiento, donde las aristas son ponderadas en función de su centralidad [11].

## 3.6 Resumen

En la **Figura 3-12** se muestra un resumen de los diferentes métodos para la detección de comunidades presentados en este capítulo.

Esta primera clasificación la utilizaremos para evaluar los algoritmos analizados y para motivar un método de evaluación adecuado.

Debido a la variedad de algoritmos presentes en la literatura que se han propuesto para dar solución a un problema general de la detección de comunidades en redes, es pertinente establecer una comparación de estos basándose en sus características principales, que consideramos que son: complejidad computacional, aplicación a problemas con solapamiento (S), soporte para redes dirigidas (D), y posibilidad de trabajar con redes ponderadas (P).

**Figura 3-12:** Clasificación de los métodos de detección de comunidades.

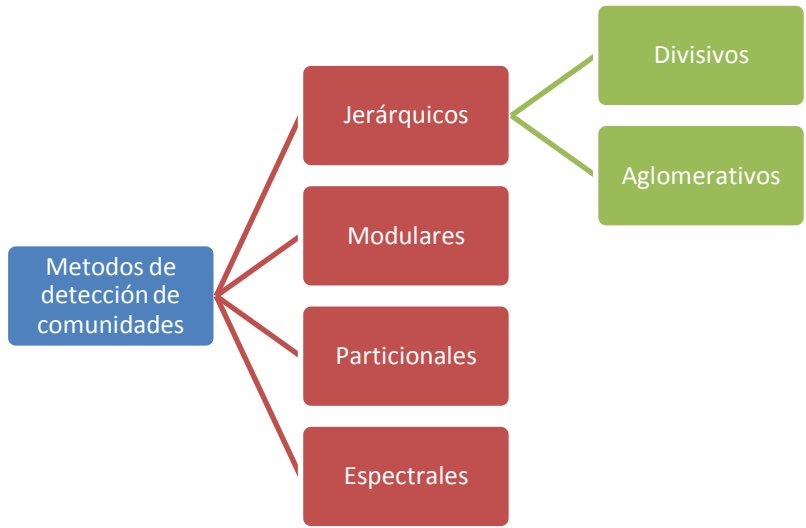

Esta clasificación se puede ver de forma resumida en la tabla que se encuentra a continuación:

**Tabla 3-1**: Comparación de la complejidad de los algoritmos analizados

| Método | Algoritmo | Año | Ref. | S | D | P | Complejidad |
|---|---|---|---|---|---|---|---|
| Jerárquico | Newman y Girvan | 2002 | [25] | | | | $O(n^3)$ |
| | Radicchi | 2004 | [61] | | X | X | $O(n^2)$ |
| | WERN-Kpath | 2012 | [12] | | X | X | $O(km)$ |
| | PAM | 2013 | [90] | X | X | X | $O(k(n-k)^2)$ |
| | SLINK | 1971 | [73] | | | X | $O(n^2)$ |
| | CLINK | 1976 | [13] | | | X | $O(n^2 \log n)$ |
| | UPGMA | 1958 | [39] | | | X | $O(n^2 \log n)$ |
| | FAC-EG | 2008 | [44] | | | X | $O(m)$ |
| | HC-PIN | 2011 | [31] | | | X | $O(\bar{k}^2 m)$ |
| | Zhao - Zhang | 2011 | [91] | X | | X | $O(n^2 \log n)$ |
| | EAGLE | 2009 | [71] | X | X | X | $O(n^2 + (h+n)s) + O(n^2 s)$ |
| Modular | Fast Greedy | 2004 | [50] | | | X | $O(n^2)$ |
| | Clauset et al. | 2004 | [7] | | | X | $O(m \log^2 n)$ |
| | RG | 2010 | [56] | | X | X | $O(m+n)$ |
| | FPMQA | 2013 | [6] | | | X | $O((k^{\max})^2)$ |
| Particional | k-medias | 1967 | [40] | | | X | $O(tkn)$ |
| | Kernighan – Lin | 1970 | [33] | | | | $O(n^2 \log n)$ |
| | LMAP | 2010 | [69] | | X | X | $O(n^3 \log n)$ |
| Espectral | Espectral Estándar | 1973 | [15] | | | X | $O(n^3)$ |
| | GANC | 2011 | [57] | | | X | $O(n \log^2 n)$ |
| Solapamiento | CPM | 2005 | [57] | X | | | $O(n_c^2)$ |
| | SCPM | 2008 | [35] | X | | X | $O(n_c)$ |
| | MOSES | 2010 | [41] | X | | | $O(n)$ |
| | Shang et al. | 2010 | [70] | X | | X | $O(n^2)$ |
| | CONA | 2011 | [87] | X | | | $O(tn_c n^2)$ |

Leyenda: ($n$)número de nodos en la red, ($m$) número de aristas, ($\bar{k}$) es el número promedio de vecinos para todos los nodos ($k^{\max}$): es el número máximo de vecinos para todos los nodos, ($t$) número de iteraciones, ($n_c$) número de *cliques,* ($s$) número máximo de *cliques* al inicio del algoritmo, ($h$)número máximo cliques vecinos conectados

Según la revisión, uno de los métodos con buenos resultados es el desarrollado por Girvan y Newman basado en coeficiente de centralidad por intermediación (*betweenness*) pero presenta varios inconvenientes como fallos en redes altamente densas y tiempos de ejecución altos por su alta complejidad computacional. El método de Radicchi se basa en el coeficiente de agrupamiento, el cual es más rápido, pero es más susceptible a fallos cuando el coeficiente de agrupamiento obtiene valores pequeños. Por sus características, se enfoca al análisis de redes densas, con gran número de triángulos.

En los métodos divisivos hay gran interés en el desarrollo de medidas de centralidad y agrupamiento, así como en la implementación de técnicas heurísticas que permitan disminuir la complejidad de los métodos, lo que permitiría su aplicación en tiempos razonables sobre grandes conjuntos de datos.

Los métodos divisivos trabajan de manera inversa a los métodos modulares, empleando medidas de similitud, con una complejidad computacional razonable del orden de $O(n^2)$. Una de sus debilidades es su tendencia a ignorar los nodos aislados o regiones poco conectadas de la red en el análisis que realizan para la detección de comunidades.

Una de las buenas alternativas para encontrar una solución de manera rápida es el uso de métodos modulares, basados en medidas de modularidad y la optimización de funciones. Se investiga la mejora de las funciones que permiten medir la modularidad de una red, para así obtener mejores resultados sin que esto suponga un detrimento en su rendimiento.

Con respecto a los métodos particionales, presentan tiempos de resolución reducidos, si bien una de sus desventajas es que se decanta por los grupos locales.

En los métodos espectrales se obtienen buenos resultados, pero con la desventaja de una alta complejidad computacional. En torno a este método, varios investigadores emplean diferentes técnicas para determinar el corte más adecuado, estudian la mejor manera de representar el resultado, o cómo mejorar su rendimiento sin disminuir la calidad de los agrupamientos.

Dados los buenos resultados que ofrecen los métodos espectrales, un área de particular interés es la investigación de cómo optimizar su rendimiento. Una de las técnicas que toma cada vez más fuerza es la paralelización de estos métodos, permitiendo obtener menores tiempos de ejecución.

Los métodos de detección de comunidades con solapamiento afrontan el problema de detectar comunidades donde un nodo puede pertenecer a más de un *cluster*. Varios de estos métodos se basan en detectar grupos altamente interconectados como los *cliques*, pero con el inconveniente de su alta complejidad computacional. Varias investigaciones se centran en desarrollar métodos menos costosos.

# 4. Evaluación de algoritmos

Una de las áreas importantes en el proceso de detección de comunidades consiste en como determinar la calidad de los *clusters*, si en realidad estos agrupamientos representan adecuadamente el comportamiento de los individuos en la red.

El objetivo principal de estas medidas de evaluación es obtener, para un *cluster* $C$, una respuesta a la pregunta ¿cómo puedo cuantificar la calidad del *cluster*?. No hay un consenso actualmente para responder esta pregunta, por consiguiente, se busca establecer criterios a través de los cuales pueda ser evaluada la calidad de los resultados, dependiendo las características de la red.

A pesar de la multiplicidad de métodos para detección de comunidades, no todos los métodos logran cumplir su objetivo, debido a que no logran cubrir todas las características de una comunidad, con lo cual surgen las siguientes preguntas [82]:

- ¿El método es sensible a pequeñas perturbaciones, como por ejemplo, el ruido?
- ¿El método es sensible al orden de los datos?
- Si se aplican dos algoritmos de agrupamiento sobre una misma red, ¿los resultados son equiparables?
- Si hay acceso a la respuesta esperada ¿el resultado obtenido se acerca a la respuesta ideal?

En la actualidad, existen varias formas de definir un buen agrupamiento. A continuación, se realizará un resumen de las principales métricas empleadas y las características resaltada por cada una de ellas. Debido a que cada métrica resalta una cualidad en particular y, como se vio anteriormente, cada método hace énfasis en una característica de las redes para la formación de los *clusters*, la respuesta a las preguntas anteriores será a menudo bastante subjetiva.

La evaluación de la calidad de un *cluster* depende del tipo de red que se está analizando. Hay que distinguir el origen de los datos, además de la existencia de información adicional que permita contrastar los resultados o los métodos con los cuales se realizó el agrupamiento. A grandes rasgos, se ha llegado a una categorización para la evaluación de la calidad de un *cluster* como las medidas se clasifican en medidas no supervisadas y medidas supervisadas [79]:

## 4.1 Evaluación no supervisada

La evaluación no supervisada se emplea en redes para las que no se tiene información externa, aparte de la que ya contiene la misma red. En este caso, se suele emplear medidas de distancia, las cuales pueden ser analizadas desde dentro del *cluster* o desde fuera.

### 4.1.1 Cohesión y separación

Una de las medidas de evaluación no supervisada es encontrar los índices de cohesión y separación entre los nodos de un *cluster* [79].

La **cohesión** se define como la suma las distancias asociadas a las aristas que conectan a los nodos dentro del *cluster*. Se espera que su resultado sea el más pequeño posible [79].

$$\text{cohesión}(C_i) = \sum_{u,v \,\in C_i} \text{proximidad}(u,v)$$

De manera similar, tenemos la **separación.** Está medida consiste en determinar la distancia a la cual están los nodos de los *clusters* con respecto a aquellos que no pertenecen al mismo. Se mide de manera similar que la cohesión, pero en este caso se espera que su resultado sea el máximo posible [79].

$$\text{separación}(C_i, C_j) = \sum_{\substack{u \in C_i \\ v \in C_j}} \text{proximidad}(u,v)$$

Pero las distancias pueden establecerse no sólo entre nodos. Podemos tener en cuenta los centroides como puntos de referencia para medir la cohesión y/o separación de los nodos o de los *clusters;* siendo $c_i$ el centroide del *cluster* $C_i$ [79]:

$$\text{cohesión}(C_i) = \sum_{u,v \in C_i} \text{proximidad}(u, c_i)$$

$$\text{separación}(C_i, C_j) = \sum_{\substack{u \in C_i \\ v \in C_j}} \text{proximidad}(u, v)$$

Las medidas de proximidad pueden ser clasificadas en dos agrupaciones principalmente: métricas de distancia y de correlación [51].

**Tabla 4-1**: Medidas de distancia

| | Medidas de distancia |
|---|---|
| Minkowski | $d_r(u,v) = \left(\sum_{k=1}^{n} |u_k - y_k|^r\right)^{\frac{1}{r}}, r \geq 1$ |
| Mahattan ($r = 1$) | $d_1(u,v) = \sum_{k=1}^{n} |u_k - y_k|$ |
| Euclídea ($r = 2$) | $d_2(u,v) = \sqrt{\sum_{k=1}^{n} (u_k - y_k)^2}$ |
| Chebychev ($r \to \infty$) | $d_{\max}(u,v) = \lim_{r \to \infty} \left(\sum_{k=1}^{n} |u_k - y_k|^r\right)^{\frac{1}{r}}$ |

**Tabla 4-2**: Medidas de correlación

| | Medidas de correlación |
|---|---|
| Escalar | $s(u,v) = u \cdot v = \sum_{k=1}^{n} u_k v_k$ |
| Coseno | $s_{\cos}(u,v) = \dfrac{u \cdot v}{\|u\|\|v\|}, \quad u = \sqrt{u \cdot u}$ |
| Pearson | $Corr(u,v) = \dfrac{n \sum u_k v_k - \sum u_k \sum v_k}{\sqrt{n \sum u_k^2 - (\sum u_k)^2} - \sqrt{n \sum v_k^2 - (\sum v_k)^2}}$ |

## 4.1.2 Evaluación individual de *clusters* y nodos

### 4.1.2.1 Coeficiente de silueta

Un método popular es el cálculo del **coeficiente de silueta,** propuesto por Rousseeuw en 1987 [68], el cual combina los conceptos de cohesión y separación [79].

De manera formal, se define como:

$$s(v_i) = \frac{b(v_i) - a(v_i)}{\max(a(v_i), b(v_i))}$$

Donde $a(v_i)$ es la distancia promedio del nodo $v_i$ a los demás nodos de su *cluster* y $b(v_i)$ es la distancia mínima que existe entre el nodo analizado y un *cluster* al cual no pertenece [79].

El coeficiente de silueta puede variar entre -1 y 1. Los valores positivos nos indican que el nodo corresponde al *cluster* en el cual se encuentra; valores negativos indican que el nodo debería ser asignado al *cluster* vecino más cercano [79].

Para un *cluster*, su coeficiente de silueta se define como [3]:

$$s(C_j) = \frac{1}{m} \sum_{i=1}^{m} s(v_i)$$

Y, de manera similar, se puede definir el coeficiente de silueta para todo un grafo [3]:

$$s(G) = \frac{1}{c} \sum_{j=1}^{c} s(C_j)$$

Una de sus grandes limitaciones es su complejidad a la hora de calcularlo, debido a que tiene que buscar todos los caminos simples entre todos los pares de nodos. Otra deficiencia es la presencia de nodos aislados o nodos hoja, que hace que los *clusters* con varios nodos hoja tengan grandes valores de coeficiente de silueta, sin importar la calidad de los demás *clusters* [1].

### 4.1.2.2 Conductancia

La **conductancia** de un *cluster* [32] permite medir la probabilidad que existe de un camino aleatorio dé un paso por el cual se pueda salir del *cluster*. Sea cut($C$) el tamaño de el corte entre $C$ y $\bar{C}$. vol($C$) es la suma de los grados de los nodos en $C$. De manera formal se define como [26]:

$$\varphi(C_i) = \frac{\text{cut}(C_i)}{\min(\text{vol}(C_i), \text{vol}(\bar{C_i}))}$$

La medida de conductancia en un grafo completo es el valor mínimo de conductancia que hay entre todos sus *clusters*[32]*:*

$$\varphi(G) = \min(\varphi(C_i)), C_i \subseteq V$$

A partir de la medida de conductancia, es posible definir las medidas de **conductancia intra-cluster** y **conductancia inter-cluster** [1].

- **Conductancia intra-cluster**

La conductancia intra-cluster es el valor mínimo de la conductancia del grafo que incide sobre el *cluster* $C_i$. Si los valores son bajos, quiere decir que el cluster puede ser excesivamente grande [1].

$$\alpha(C) = \min \varphi(G[C_i]), \ i \in \{1, \dots, k\}$$

- **Conductancia inter-cluster**

La conductancia inter-cluster es el complemento del valor máximo de la conductancia del *cluster*,

$$\sigma(C) = 1 - \max \varphi(C_i), \ i \in \{1, \dots, k\}$$

donde los valores entre más bajos indicarán que al menos uno de los *clusters* tiene fuertes conexiones fuera de él [1].

La minimización de la conductancia para todos los cortes en un grafo y buscar el corte adecuado es un problema NP-duro [5].

Tanto la conductancia interna, como la externa se miden en el intervalo de [0,1]. Para determinar un buen *cluster*, los valores de conductancia (interna y externa) tienen que ser altos [5]. Una de las ventajas del método de conductancia es que funciona muy bien para agrupamientos con poca cantidad de *clusters*. Cuando se tiene una mayor cantidad de *clusters* de menor tamaño, estos tienden a tener más aristas de enlace entre ellos, por lo que tiende a fallar la métrica [1].

### 4.1.2.3 Cobertura

La **cobertura** es la relación que existe entre el número o el peso de las aristas de un *cluster* y el peso de todo el grafo. De manera formal, se define como [5]:

$$cov(C_i) = \frac{w(C_i)}{w(G)}$$

De manera intuitiva, cuanto mayor sea el valor de cobertura en el *cluster*, mejor será la calidad del mismo. Esta conclusión es similar a la desarrollada en el corte mínimo (*mincut*) [5]. Una desventaja, al igual que la medida de conductancia, es su tendencia a calificar de mejor manera a los *clusters* de gran tamaño [1].

### 4.1.2.4 Rendimiento

El rendimiento es una medida que permite contar la fracción de aristas dentro del *cluster* con pares de nodos no adyacentes entre nodos del *cluster* y nodos de los demás *clusters*. De manera formal, se define como [5]:

$$perf(C) = 1 - \frac{2m(1 - 2cov(C)) + \sum_{i=1}^{k}|C_i|(|C_i| - 1)}{n(n-1)}$$

donde

$$m = w(G), \ n = |V|$$

El valor del rendimiento se mide en el intervalo [0,1], donde los valores más altos indican que un *cluster* es denso internamente y con pocas conexiones de manera externa, lo que da a entender que es un buen *cluster*. Pero en redes de gran tamaño, no obstante, hay tendencia a devolver resultados de gran valor aparente por la gran cantidad de *clusters* en la red [1].

### 4.1.2.5 Agrupamiento o *clustering*

Una de las medidas más comunes para medir la calidad de los *clusters* es el coeficiente de agrupamiento (ver sección 2.4.4), la cual evalúa el nivel de agrupamiento de los nodos. Se fundamenta en medir la cantidad de triángulos presentes en el *cluster*.

Recordando, el coeficiente de agrupamiento para un *cluster* es [64]:

$$C_i = \frac{2m_i}{k_i(k_i - 1)}$$

donde $m_i$ es el número de conexiones entre los $k_i$ vecinos del nodo $i$.

El coeficiente de agrupamiento para un grafo es [48].

$$C_g = \frac{1}{n}\sum_{i \in V} C_i$$

El rango del coeficiente de agrupamiento está comprendido en el intervalo de [0,1], donde los valores más altos del coeficiente indican un "buen" agrupamiento.

### 4.1.2.6 Modularidad

El índice o **coeficiente de modularidad**, propuesto por Girvan y Newman [52], es una de las métricas ampliamente utilizadas por métodos para la detección de comunidades. Está estrechamente relacionada con la medida de cobertura, pero supera algunas de sus deficiencias. Como se explico anteriormente (ver sección 2.4.5) se fundamenta en comparar el conjunto de enlaces dentro del *cluster* analizado frente a si ese conjunto fuera construido de manera aleatoria.

Se fundamenta en la determinación de las densidades dentro y fuera de cada *cluster*, que son comparados con la densidad generar que tendría la red si fuera construida de manera aleatoria [51].

De manera formal, el coeficiente de modularidad se define como:

$$Q = \sum_r (e_{rr} - a_r^2)$$

donde $e_{rr}$ son el conjunto de aristas que se encuentran en un *cluster* $r$ y $a_r$ representa el conjunto de nodos que forman parte del *cluster* $r$ [51].

Los valores de modularidad cercanos a cero indican que el número de enlaces dentro del *cluster* es menor al valor que se esperaría en una red aleatoria. El valor máximo de está métrica tiende a 1. Valores muy cercanos indican que el *cluster* está altamente conectado. De manera práctica, se ha observado que las comunidades presentan un valor de modularidad entre 0.3 a 0.7 [52].

Una debilidad de está métrica es la tendencia a fallar cuando la red tiene muchos nodos aislados. Al no tener más integrantes en su interior, su nivel de conexión es 0. Si hay muchos grupos, estos tienen a eclipsar las demás agrupaciones, provocando un nivel de modularidad bajo [1].

## 4.2 Evaluación supervisada por similitud

La evaluación supervisada consiste en medir la calidad de los *cluster*s con información externa a la encontrada en la red analizada. En este tipo de evaluación hay dos enfoques: **por clasificación** y **por similitud** [79].

La comprobación de rendimiento de un algoritmo necesita establecer un criterio de "similitud". En este caso, se establece una medida mediante la cual dos resultados de un proceso de *clustering* sobre una misma red puedan evaluarse directamente con un índice de similitud [19].

Sea $X = (X_1, X_2, \ldots, X_{nX})$ y $Y = (Y_1, Y_2, \ldots, Y_{nY})$ dos conjuntos de particiones genéricas de un grafo $G$ con $n_X$ y $n_Y$ *clusters*, respectivamente. Siendo $n_i^X$ y $n_j^Y$ el número de nodos en el *cluster* $X_i$ y $Y_j$, respectivamente, y la matriz de confusión $n_{ij}$ el número de nodos compartidos por los *clusters* $X_i$ y $Y_j$ [19]:

$$n_{ij} = |X_i \cap Y_j|, \quad 1 \leq i \leq n_X, \quad 1 \leq j \leq n_Y$$

Las medidas de similitud evalúan el número de pares de nodos que se encuentran bajo el mismo *cluster* encontrado por dos métodos diferentes sobre una misma red, se definen de la siguiente forma:

- $a_{11}$ indica que el número de de pares de nodos que son del mismo *cluster* en ambas particiones.
- $a_{01}$ indica que el número de pares de nodos que están en un mismo *cluster* en $X$ y diferentes en el *cluster* $Y$.
- $a_{10}$ indica que el número de pares de nodos que están en un mismo *cluster* en $Y$ y diferentes en el *cluster* $X$.
- $a_{00}$ indica que el par de nodos evaluados se encuentran en diferentes *clusters* en ambos conjuntos de particiones.

Si se tiene conocimiento previo de la comunidad evaluada, ésta se puede utilizar como punto de comparación, midiendo el resultado del algoritmo frente a un resultado de referencia al estilo de las técnicas de aprendizaje supervisado.

La evaluación supervisada de un proceso de *clustering*, según Fortunato [19], puede clasificarse en 3 tipos de medidas: comparación de pares, concordancia de *cluster* y teoría de la información.

## 4.2.1 Comparación de pares

### 4.2.1.1 Coeficiente de Wallace

El coeficiente de Wallace [83] propone en 1983 dos medidas de similitud:

$$\mathcal{W}_{XY} = \frac{a_{11}}{\sum_k n_k^x(n_k^x - 1)/2} \qquad \mathcal{W}_{YX} = \frac{a_{11}}{\sum_k n_k^y(n_k^y - 1)/2}$$

donde $W_{XY}$ y $W_{YX}$ representan la probabilidad de que un par de nodos de un mismo *cluster* de $X$ se encuentren en el mismo *cluster* que en $Y$, y viceversa. A pesar de evaluar el mismo conjunto de nodos, los coeficientes son asimétricos [19].

Estas medidas también son conocidas como precisión (*precisión*) y recuperación (*recall*), las cuales son base para otro tipo de medidas como la medida – F (*f-measure*) [39].

#### 4.2.1.2 Coeficiente de Rand

El coeficiente de Rand [62] se basa en comprobar el resultado de un algoritmo de agrupamiento frente a un resultado correcto previamente conocido.

De manera general, calcula el porcentaje de nodos correctamente clasificados entre todo el conjunto de elementos. De manera natural, Rand extendió el concepto plateado en la medidas de rendimiento [19].

El coeficiente de Rand se define como:

$$\mathcal{R}(X,Y) = \frac{a_{11} + a_{00}}{a_{11} + a_{01} + a_{10} + a_{00}} = \frac{2(a_{11} + a_{00})}{n(n-1)}$$

El valor del coeficiente está entre 0 y 1, donde 0 indica que no hay ningún par de nodos clasificado de la misma forma en ambos métodos y, recíprocamente, 1 indica que todos los pares de nodos tienen la misma clasificación en ambos métodos. Una de las limitaciones del coeficiente de Rand es su alta dependencia al número de *clusters* analizados [82].

#### 4.2.1.3 Coeficiente de Mirkin

El coeficiente de Mirkin, también conocido como *equivalence mismatch distance*, es una variación del coeficiente de Rand. Se define en los siguientes términos[19]:

$$\mathcal{M}(X,Y) = 2(a_{01} + a_{10}) = \frac{n(n-1)}{[1 - \mathcal{R}(X,Y)]}$$

Es una medida muy sensible con respecto al tamaño de los *clusters*. Si se analizan dos *clusters* con la misma cantidad de elementos, puede tener un valor más alto que aplicado a dos *clusters* en los cuales uno sea una derivación del otro [82].

#### 4.2.1.4 Coeficiente de Jaccard

El coeficiente de Jaccard permite medir la relación entre el número de pares de nodos clasificados en el mismo *cluster* en ambas particiones y el número de pares de nodos en el mismo *cluster* en, al menos, una partición [19].

Para *clusters* no solapados, se define como [63]:

$$\mathcal{J}(X,Y) = \frac{|X \cap Y|}{|X \cup Y|}$$

Si se despeja el coeficiente de Jaccard en términos de la cantidad de nodos evaluados, la expresión queda de la siguiente forma [19]:

$$\mathcal{J}(X,Y) = \frac{a_{11}}{a_{11} + a_{01} + a_{00}}$$

## 4.2.2 Concordancia de *cluster* y solapamiento

Hay otro tipo de medidas que permiten detectar el nivel de solapamiento de los *clusters*. A continuación se revisarán algunas de estas medidas. Varias de éstas presentan asimetrías, lo que puede dificultar su uso en la práctica:

### 4.2.2.1 Coeficiente de solapamiento de Jaccard

Uno de los inconvenientes del coeficiente de Jaccard es que tiende a fallar cuando los dos conjuntos tienen una gran diferencia de tamaño. En este caso, es mejor emplear el coeficiente de solapamiento [63].

El coeficiente con solapamiento es [63]:

$$O(X,Y) = \frac{|X \cap Y|}{\min(|X|,|Y|)}$$

Si este coeficiente tiene un valor de 1, esto indica que el conjunto más pequeño es un subconjunto del grande. Si, en cambio, se obtiene un valor de 0, los dos conjuntos no tienen nodos comunes [63].

### 4.2.2.2 Medida-F (*F-measure*)

La medida-F, o *f-measure* en inglés (también como *f-score* [39]) permite medir el grado en que un *cluster* contiene nodos de un *cluster* base. Esta similitud corresponde al resultado de media armónica de las medidas de precisión y recuperación (*recall*) [82], típicas de los sistemas de recuperación de información:

$$\mathcal{F}(X,Y) = \frac{2|X||Y|}{|X|+|Y|} = \frac{2r_{ij}p_{ij}}{r_{ij}+p_{ij}}$$

donde la precisión se define como:

$$p_{ij} = \frac{a_{11}}{a_{11}+a_{01}}$$

y la recuperación como:

$$p_{ij} = \frac{a_{11}}{a_{11}+a_{10}}$$

Al ser una medida asimétrica, es ideal para comparar el *cluster* con una solución óptima [82].

### 4.2.2.3  Coeficiente de Meila – Heckerman

El coeficiente de Meila – Heckerman [43] emplea el mismo principio que la medida-f. Permite comparar el resultado de un método de agrupamiento con una solución óptima conocida [19]:

$$\mathcal{MH}(X,Y) = \frac{1}{n}\sum_{i=1}^{n_X} \max_{Y_j \in X} n_{ij}$$

donde $X$ es el agrupamiento del método a evaluar e $Y$ es el agrupamiento óptimo. Al ser asimétrica la comparación, resulta ideal para comparar un resultado frente a un óptimo. En base a esta medida se elaboró una versión simétrica llamada *Maximun-Match-Measure* o medida de máximo emparejamiento [82].

### 4.2.2.4  *Maximum-Match-Measure*

Basado en el coeficiente de Meila – Heckerman, permite evaluar dos técnicas de agrupamiento. Al ser una medida simétrica, se limita a las entradas de la matriz de confusión donde coincidan los conjuntos de *clusters* comprados [82].

$$\mathcal{MM}(X,Y) = \frac{1}{n}\sum_{i=1}^{\min(n_X,n_Y)} n_{ii\prime}$$

#### 4.2.2.5  Coeficiente de Van Dongen

El coeficiente de Van Dongen [80] es una medida simétrica para la comparación de dos métodos de agrupamiento. Se basa en encontrar la máxima cantidad de intersecciones entre los *clusters* analizados[19]:

$$\mathcal{D}(X,Y) = 1 - \frac{1}{2n}\left[\sum_{k=1}^{n_X} \max_{k'} n_{kk'} + \sum_{k'=1}^{n_Y} \max_{k} n_{kk'}\right]$$

### 4.2.3  Teoría de la Información

Un tercer grupo de tipos de medida de similitud son aquellas basadas en la Teoría de la Información. Dos particiones son similares si se necesita poca información para inferir que una partición sea igual a otra. La cantidad de información adicional que se necesita se utiliza como medida de similitud. Esta cantidad de información se suele medir por la entropía [19].

#### *4.2.3.1  Entropía de un cluster*

La entropía es una medida que permite evaluar el grado de "desorden" que presenta un sistema. De manera más formal, es la suma ponderada de las cantidades de información de todos los posibles estados del agrupamiento $C$ [82]:

$$\mathcal{H}(C) = -\sum_{i=1}^{k} P(i) \log_2 P(i)$$

donde para cada *cluster* $C_i$ es calculado la probabilidad

$$P(i) = \frac{|C_i|}{n}, \qquad C_i \in C$$

Entre menor sea la entropía, más estable es el agrupamiento [19].

#### 4.2.3.2 Información mutua

A partir de la definición de entropía, se define el concepto de información mutua. Esta medida indica en cuánto se podría reducir la incertidumbre de un nodo aleatorio sobre un *cluster* si se conoce de antemano el estado de ese nodo y su *cluster* pertinente bajo otro agrupamiento [82].

$$\mathcal{I}(X,Y) = \sum_{i=1}^{n_X}\sum_{j=1}^{n_Y} P(i,j) \log_2 \frac{P(i,j)}{P(i)P(j)}$$

donde $P(i,j)$ es la probabilidad de que un nodo pertenezca a un *cluster* $X_i$ en el agrupamiento $X$ y que también pertenezca al *cluster* $Y_j$ en el agrupamiento $Y$ [82].

$$P(i,j) = \frac{|X_i \cap Y_j|}{n}$$

El inconveniente de esta medida es que no está limitada por un valor constante, lo que hace difícil su interpretación, por lo que es conveniente realizar un proceso de normalización [19].

#### 4.2.3.3 Información mutua normalizada

Una forma de dar solución al problema de la medida de información mutua es la implementación de una normalización para limitar la entropía de las medidas. Esta normalización puede ser aritmética o geométrica [82].

La normalización aritmética [20]:

$$\mathcal{NMI}_2(X,Y) = \frac{2\mathcal{I}(X,Y)}{\mathcal{H}(X) + \mathcal{H}(Y)}$$

La normalización geométrica [76]:

$$\mathcal{NMI}_1(X,Y) = \frac{\mathcal{I}(X,Y)}{\sqrt{\mathcal{H}(X)\mathcal{H}(Y)}}$$

Este tipo de tratamiento de la información permite analizar los *clusters* sin necesidad de conocer directamente el método por el cual se realizó el agrupamiento o las características originales de la red. El rango posible de medida de $\mathcal{NMI}_1(X,Y)$ viene dado de 0 a 1, donde $\mathcal{NMI}_1(X,Y) = 1$ si las particiones son idénticas y $\mathcal{NMI}_1(X,Y) = 0$ cuando las particiones son independientes [82].

Una de las limitaciones de este tipo de medida es su implementación únicamente sobre agrupamientos sin solapamiento. Aunque hay variantes para el análisis de solapamientos como la propuesta por Lancichinetti [36], no son del todo estables [82].

#### 4.2.3.4 Variación de la información

También hay otras variaciones, entre estas destaca la propuesta por Meilă [42]:

$$\begin{aligned} \mathcal{VI}(X,Y) &= \mathcal{H}(X|Y) + \mathcal{H}(Y|X) \\ &= \big(\mathcal{H}(X) - \mathcal{I}(X,Y)\big) + \big(\mathcal{H}(Y) - \mathcal{I}(X,Y)\big) \end{aligned}$$

Esta medida se compone de la cantidad de información de $X$ que se ha perdido junto con la información sobre $Y$ que todavía es posible ganar al pasar de la agrupación de $X$ a $Y$ [82].

## 4.3 Formalización

Como se ha podido observar, uno de los grandes inconvenientes de cara a la evaluación de los algoritmos de detección de comunidades es encontrar el procedimiento o los parámetros adecuados para realizar este procedimiento. Esta situación da como resultado una evaluación precaria de la eficacia de un método, de detección de comunidades.

Uno de los trabajos de Kleinberg [34] demuestra la complejidad del problema del *clustering*. De manera formal, propone 3 axiomas que debería cumplir un algoritmo de clustering, demostrando su imposibilidad.

- Invariancia de la escala: El cambio en la escala de la medida de distancia no debe provocar un cambio en los resultados de los algoritmos.

- Riqueza: El método debe tener la posibilidad de contemplar todas las posibles soluciones de partición en un problema de *clustering.*

- Consistencia: No deben cambiar los resultados de un algoritmo si hay cambios en las medidas intra-cluster e inter-cluster.

Hay variaciones que flexibilizan los axiomas para el desarrollo de algoritmos que puedan cumplir estos tres axiomas, una vez relajados. [1]

También existen trabajos en los cuales se realiza comparaciones empíricas de los métodos de detección de comunidades con varias métricas, como los realizados por Danon *et al.* [10], Fortunato [19], Lancichinetti *et al.* [37], Meila *et al.* [42].

Estos trabajos intentan exponer la mejor forma de comparar adecuadamente los algoritmos, ver desde un punto de vista comparativo que parámetros deben tenerse en cuenta en el momento de escoger el método adecuado para un problema en particular, y qué características de los *clusters* se buscan de forma específica, (por ejemplo densidad, centralidad, modularidad, numero de clusters o eficiencia).

Esta problemática sugiere las siguientes preguntas: ¿cómo definir un *cluster* como "bueno"?, ¿qué parámetros debe contar un método para que pueda ser calificado como "bueno"?. Trabajos como el presentado por Wagner [82] hacen una recopilación de las características que debe tener en cuenta un agrupamiento para considerarse "bueno", formulando la siguiente definición:

Sea $f$ una medida para comparar los agrupamientos de un conjunto $X$ y sean $C, C', C'' \in P(X)$ agrupamientos de $X$. $f$ debe tener las siguientes características:

1. Puede ser aplicado a todos los clusters de $P(X)$.
2. No debe tener restricción con respecto a la estructura de los agrupamientos como:
    a. Tamaño de los *clusters*.
    b. Número de *clusters* ($C, C'$ pueden tener diferentes cantidades de *clusters*).
    c. Dependencias entre $C$ y $C'$.
3. Independencia en el número de *clusters*.
4. Independencia en el número de elementos.

## 4.4 Resumen

En este capítulo se han revisado diferentes formas de evaluar la calidad de un método de agrupamiento. Estas formas se dividen en función de la información adicional de la que se disponga para su evaluación.

**Figura 4-1:** Clasificación de los métodos de evaluación de comunidades.

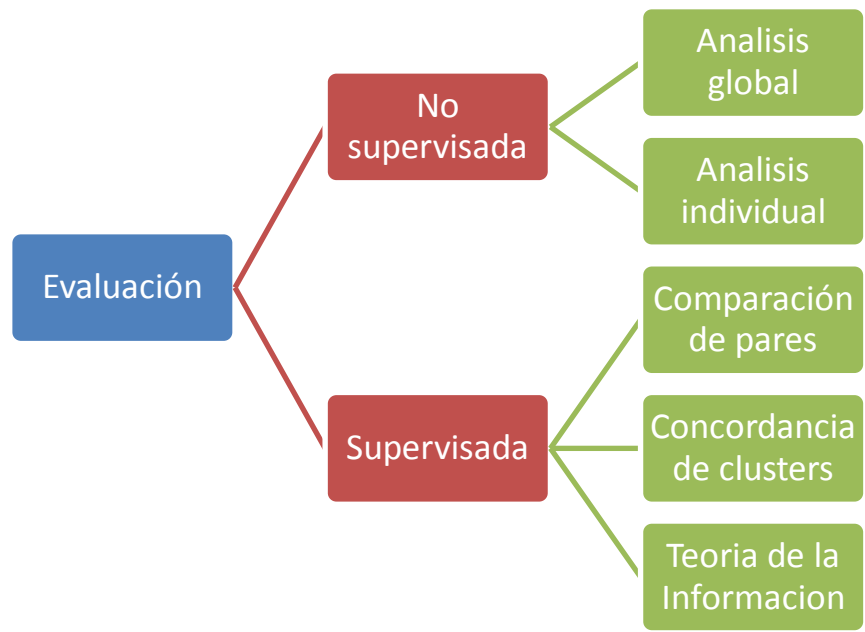

Los métodos no supervisados son aquellos en los cuales no se dispone de información adicional para su análisis, sólo la contenida en la red y en el resultado del método de agrupamiento empleado. En su mayoría, los métodos empleados se basan en medidas de similitud entre nodos y *clusters*. Algunos ejemplos representativos de ellos son el coeficiente de silueta, la conductancia o la modularidad.

Los métodos basados en similitud utilizan información externa a la contenida en la red y el resultado del método. Las medidas se pueden dividir de varias formas, una de ellas es por su simetría a la hora de evaluar. En caso de ser simétricas, se evalúa la similitud del *cluster* frente al agrupamiento utilizado en otro método sobre la misma red. En las mediciones asimétricas, se compara el método con respecto a una respuesta deseada.

A continuación se resume las medidas en dos tablas clasificatorias. La primera tabla (**Tabla 4-3**) agrupa las medidas no supervisadas, teniendo en cuenta la referencia de la medida, el rango de la medida y sus características principales. Cada medida tiene sus fortalezas y debilidades.

**Tabla 4-3**: Características principales de los métodos no supervisados de evaluación de un agrupamiento

| No supervisado | | | | |
|---|---|---|---|---|
| Enfoque | Medida | Ref. | Rango | Características |
| Análisis Global | Cohesión | [79] | $[0, \infty]$ | Mide las distancias entre nodos dentro de un cluster, se buscan valores pequeños, varía dependiendo de la medida de proximidad. |
| | Separación | [79] | $[0, \infty]$ | Mide las distancias de los nodos del *cluster* con respecto a aquellos que no pertenece, se busca el máximo posible, varía dependiendo de la medida de proximidad. |
| Análisis individual | Coeficiente de silueta | [68] | $[-1, 1]$ | Adecuada para comunidades altamente conectadas. Alta complejidad y fallos con nodos hoja. |
| | Conductancia | [32] | $[0,1]$ | Medición de cuellos de botella, adecuado para *clusters* de gran tamaño; fallos en la evaluación de *clusters* con pocos nodos, pequeños y/o muy grandes. |
| | Cobertura | [5] | $[0,1]$ | Peso del *cluster*, basado en los cortes mínimos; fallos en la evaluación de *clusters* con pocos nodos, pequeños y/o muy grandes. |
| | Rendimiento | [5] | $[0,1]$ | Número de nodos adyacentes, densidad; falla en redes de gran tamaño con numero alto de *clusters*. |
| | Coeficiente de agrupamiento | [64] | $[0,1]$ | Búsqueda de estructuras conexas (triángulos). |
| | Modularidad | [52] | $[0,1]$ | Comparación del cluster con estructura aleatoria. |

En la segunda tabla (**Tabla 4-4**) se realiza la comparación de las medidas supervisadas, que se basan en comparar el resultado de un método con respecto a una solución de referencia la proporcionada por otro método, lo que permite obtener un coeficiente de similitud entre ambos agrupamientos. Además de las características de la primera tabla, se incluye si la medida se utiliza para comparar dos métodos (simétrica) o para comparar resultados de referencia con una solución de referencia (asimétrica).

**Tabla 4-4**: Características principales de los métodos supervisados basados en similitud para la evaluación de agrupamiento

| Similitud | | | | | |
|---|---|---|---|---|---|
| Tipo | Nombre | Ref. | Sim. | Rango | Características |
| Comparación de pares | Wallace | [83] | X | [0,1] | Mide la probabilidad de un par de nodos con respecto a la pertenencia de un *cluster*. |
| | Rand | [62] | X | [0,1] | Depende del número de grupos analizados. |
| | Mirkin | [19] | X | [0,1] | Sensible al tamaño del *cluster*. |
| | Jaccard | [19] | X | [0,1] | Mide la probabilidad de pertenencia a un cluster con respecto a otro. |
| Solapamiento | Solapamiento | [19] | X | [0,1] | Mide el nivel de solapamiento de un cluster con respecto a otro. |
| | F-measure | [82] | | [0,1] | Mide el grado en que un *cluster* contiene datos de un *cluster* base. |
| | Meila-Heckerman | [43] | | [0,1] | Similar al f-measure. |
| | Maximum-match-measure | [82] | X | [0,1] | Caso de Meila. Heckerman pero simétrico. |
| | Van Dongen | [80] | X | [0,1] | Máxima cantidad de intersecciones entre los *clusters*. |
| Teoría de la Información | Entropía | [82] | | [0, ∞] | Medida del grado de incertidumbre sobre un cluster conformado por elementos al azar. |
| | Información mutua | [82] | | [0, ∞] | Difícil de predecir. |
| | Información mutua normalizada | [20] [76] | | [0,1] | No permite solapamiento. |
| | Variación de información | [42] | | [0,1] | Basado en el flujo de información de pérdida y ganancia entre *clusters*. |

Como se puede ver en las características de los métodos de las tablas anteriores y el trabajo realizado por Wagner [82], no existe un consenso sobre el cual se pueda determinar si una medida es apropiada o no. Deben ser analizadas las características del problema concreto que se desea resolver para seleccionar la medida correcta.

# 5. Conclusiones y trabajo futuro

En este trabajo de fin de máster se ha realizado un estudio de los distintos métodos que se han propuesto para la detección de comunidades en redes. Se ha realizado un análisis de las características de cada método, incluyendo su complejidad computacional. Así mismo, se han analizado varias métricas para evaluar la calidad de los resultados obtenidos por los métodos de detección de comunidades.

Los resultados alcanzados tras la realización de este trabajo son los siguientes:

- Se ha realizado una revisión bibliográfica de los diferentes métodos de detección de comunidades en redes que se han propuesto durante los últimos años.
- Se ha realizado una clasificación de los principales métodos de detección de comunidades de acuerdo a sus características, analizando sus fortalezas y sus debilidades, complejidad computacional incluida.
- Se ha realizado una clasificación de los principales métodos de evaluación de la calidad de las comunidades identificadas por los métodos de detección de comunidades, analizando las características de las distintas métricas de evaluación.
- Se han identificado varias líneas de investigación viables en el ámbito de las técnicas de detección de comunidades.

En el capítulo 2, se pudo comprobar cómo las técnicas que sirven de base para resolver problemas de detección de comunidades están aún en desarrollo. La complejidad computacional de la detección de *k-cliques* y *k-plexes* no permite que el problema sea resuelto de forma exacta para grandes redes, por lo que han de desarrollarse técnicas heurísticas que permitan resolver el problema en un tiempo razonable.

Además, no existe una definición única de "comunidad". La idea de comunidad puede interpretarse de distintas formas, por lo que no existe una única medida que tenga en cuenta todos los aspectos que pueden ser de interés en un contexto dado. Debido a esto, para la selección de métricas adecuadas pueden tenerse en cuenta varias características:

- Las características intrínsecas del problema, tales como el tamaño de la red (número de nodos y aristas) o su densidad.
- El número de comunidades, la disparidad en el tamaño de las comunidades o la existencia de solapamiento entre ellas.
- El método empleado para la detección de las comunidades.
- La disposición de información adicional (p.ej. métodos supervisados).
- Los resultados obtenidos por diferentes métodos.

En el capítulo 3, se revisó una amplia gama de métodos de detección de comunidades, que se agruparon en función de su enfoque. Entre los métodos propuestos existen algunos escalables, con complejidad computacional del orden de $O(n)$, pero que no ofrecen garantías (esto es, con tendencia a quedarse atascados en mínimos locales). Hay métodos más robustos, que ofrecen mejores resultados, pero que tienen una complejidad computacional muy alta, del orden de $O(n^3)$, lo cual los hace inviables para el análisis de redes de gran tamaño.

También se estudió el análisis de redes con comunidades solapadas, dado que las técnicas tradicionales no están diseñadas para enfrentarse a este tipo de problemas.

En el capítulo 4, se revisaron las distintas métricas que se han propuesto para evaluar la calidad de los resultados de un algoritmo de detección de comunidades. Estas métricas proporcionan distintas respuestas a la pregunta ¿cómo puedo cuantificar la calidad de una comunidad? Estas métricas se dividen en medidas no supervisadas y medidas supervisadas.

Las limitaciones identificadas durante la realización de este trabajo sugieren varias líneas de investigación que podrían dar lugar a futuros trabajos:

- El desarrollo de nuevos métodos de detección de comunidades para redes de gran tamaño que sean escalables. Estos métodos pueden basarse en:
    - El empleo de técnicas heurísticas que permitan disminuir la complejidad computacional de los algoritmos de detección de comunidades.

- o La combinación de métodos eficientes desde el punto de vista computacional con otras técnicas más robustas que permiten obtener resultados de mayor calidad.
- o La paralelización de los métodos robustos de detección de comunidades, aprovechando los avances en las capacidad de cómputo del hardware disponible comercialmente, que incluye microprocesadores con varios núcleos y SMT, coprocesadores gráficos (GPUs programables con CUDA u OpenCL) o entornos de computación distribuida (grids y cloud computing).
- o La adaptación de los métodos de detección de comunidades a las características del problema concreto, teniendo en cuenta tanto las características de la red analizada (p.ej. número de nodos o densidad) como información adicional proveniente de otras fuentes (p.ej. conocimiento experto).

- El desarrollo de nuevos métodos de detección de comunidades que permitan la detección de comunidades solapadas aprovechando la eficiencia de los métodos que se han propuesto para detecta comunidades no solapadas.

- La formalización de una metodología que permita la evaluación de la calidad de los resultados obtenidos por los métodos de detección de comunidades. Siguiendo las sugerencias formales planteadas por Wagner, esta metodología debería permitir la comparación de los resultados obtenidos por métodos de detección de comunidades alternativos.

# Bibliografía